\documentclass[prl,aps,twocolumn,floats,nofootinbib,superscriptaddress]{revtex4-1}
\usepackage{times}
\usepackage{amsmath,amssymb,graphicx,psfrag}
\usepackage[colorlinks,linkcolor={blue},citecolor={blue},urlcolor={blue}]{hyperref}% add hypertext capabilities

\begin{document}
\renewcommand{\ni}{{\noindent}}
\newcommand{\dprime}{{\prime\prime}}
\newcommand{\be}{\begin{equation}}
\newcommand{\ee}{\end{equation}}
\newcommand{\bea}{\begin{eqnarray}} 
\newcommand{\eea}{\end{eqnarray}}
\newcommand{\bal}{\begin{align*}}
\newcommand{\eal}{\end{align*}}
\newcommand{\la}{\langle}
\newcommand{\ra}{\rangle} 
\newcommand{\dg}{\dagger}
\newcommand\lbs{\left[}
\newcommand\rbs{\right]}
\newcommand\lbr{\left(}
\newcommand\rbr{\right)}
\newcommand\f{\frac}
\newcommand\e{\epsilon}
\newcommand\ua{\uparrow}
\newcommand\da{\downarrow}
\newcommand{\bcen}{\begin{center}}
\newcommand{\ecen}{\end{center}}
\newcommand{\btab}{\begin{tabular}}
\newcommand{\etab}{\end{tabular}}
\newcommand{\bdes}{\begin{description}}
\newcommand{\edes}{\end{description}}
\newcommand{\mc}{\multicolumn}
\newcommand{\ul}{\underline}
\newcommand{\non}{\nonumber}
\newcommand{\etal}{et.~al.\ }
\newcommand{\half}{\frac{1}{2}}
\newcommand{\bary}{\begin{array}}
\newcommand{\eary}{\end{array}}
\newcommand{\benum}{\begin{enumerate}}
\newcommand{\eenum}{\end{enumerate}}
\newcommand{\bitem}{\begin{itemize}}
\newcommand{\eitem}{\end{itemize}}
\newcommand{\cuup}[1]{c_{#1 \uparrow}}
\newcommand{\cdown}[1]{c_{#1 \downarrow}}
\newcommand{\cdup}[1]{c^\dagger_{#1 \uparrow}}
\newcommand{\cddown}[1]{c^\dagger_{#1 \downarrow}}

% bold greek characters
%
\newcommand{\beps}{\mbox{\boldmath $ \epsilon $}}
\newcommand{\bsig}{\mbox{\boldmath $ \sigma $}}
\newcommand{\bpi}{\mbox{\boldmath $ \pi $}}
\newcommand{\bkap}{\mbox{\boldmath $ \kappa $}}
\newcommand{\bgam}{\mbox{\boldmath $ \gamma $}}
\newcommand{\bphi}{\mbox{\boldmath $ \phi $}}
\newcommand{\balp}{\mbox{\boldmath $ \alpha $}}
\newcommand{\beot}{\mbox{\boldmath $ \eta $}}
\newcommand{\btau}{\mbox{\boldmath $ \tau $}}
\newcommand{\blam}{\mbox{\boldmath $ \lambda $}}
\newcommand{\bomg}{\mbox{\boldmath $ \omega $}}
\newcommand{\bOmg}{\mbox{\boldmath $ \Omega $}}
\newcommand{\bxhi}{\mbox{\boldmath $ \xi $}}
\newcommand{\bmu} {\mbox{\boldmath $ \mu $}}
\newcommand{\bnu} {\mbox{\boldmath $ \nu $}}
\newcommand{\bdelta}{{\boldsymbol{\delta}}}
\newcommand{\bTheta}{{\boldsymbol{\Theta}}}
\newcommand{\bpsi}{\mbox{\boldmath $ \psi $}}
\newcommand{\brho}{\mbox{\boldmath $ \rho $}}
\newcommand{\bGam}{\mbox{\boldmath $ \Gamma $}}
\newcommand{\bLam}{\mbox{\boldmath $ \Lambda $}}
\newcommand{\bPhi}{\mbox{\boldmath $ \Phi $}}
%
% bold latin
%
%\newcommand{\ba} { \mbox{\boldmath $a$}}
\newcommand{\ba} { \bm{a} }
\newcommand{\bb} { \mbox{\boldmath $b$}}
\newcommand{\bc} { {\mathbf c} }
\newcommand{\bd} { \mbox{\boldmath $d$}}
\newcommand{\bff}{ \mbox{\boldmath $f$}}
\newcommand{\bg} { \mbox{\boldmath $g$}}
\newcommand{\bh} { \mbox{\boldmath $h$}}
\newcommand{\bi} { \mbox{\boldmath $i$}}
\newcommand{\bj} { \mbox{\boldmath $j$}}
\newcommand{\bk} { \bm{k} }
\newcommand{\bl} { \mbox{\boldmath $l$}} 
\newcommand{\bmm} { \mbox{\boldmath $m$}}
\newcommand{\bn} { \mbox{\boldmath $n$}}
\newcommand{\bo} { \mbox{\boldmath $o$}}
\newcommand{\bp} { \bm{p} }
\newcommand{\bq} { \bm{q} }
\newcommand{\br} { \boldsymbol{r}}
\newcommand{\bs} { \mbox{\boldmath $s$}}
\newcommand{\bt} {\boldsymbol{t}} 
\newcommand{\bu} { \mbox{\boldmath $u$}}
\newcommand{\bv} { \mbox{\boldmath $v$}}
\newcommand{\bw} { \mbox{\boldmath $w$}}
\newcommand{\bx} { \mbox{\boldmath $x$}}
\newcommand{\by} { \mbox{\boldmath $y$}}
\newcommand{\bz} { \mbox{\boldmath $z$}}
\newcommand{\bA} { \mbox{\boldmath $A$}}
\newcommand{\bB} { \mbox{\boldmath $B$}}
\newcommand{\bC} { \mbox{\boldmath $C$}}
\newcommand{\bD} { \mbox{\boldmath $D$}}
\newcommand{\bF} { \mbox{\boldmath $F$}}
\newcommand{\bG} { \mbox{\boldmath $G$}}
\newcommand{\bH} { \mbox{\boldmath $H$}}
\newcommand{\bI} { \mbox{\boldmath $I$}}
\newcommand{\bJ} { \mbox{\boldmath $J$}}
\newcommand{\bK} { \mbox{\boldmath $K$}}
\newcommand{\bL} { \mbox{\boldmath $L$}}
\newcommand{\bM} { \mbox{\boldmath $M$}}
\newcommand{\bN} { \mbox{\boldmath $N$}}
\newcommand{\bO} { \mbox{\boldmath $O$}}
\newcommand{\bP} { \mbox{\boldmath $P$}}
\newcommand{\bQ} { \boldsymbol{Q} }
\newcommand{\bR} { {\mathbf R} }
\newcommand{\bS} { \mbox{\boldmath $S$}}
\newcommand{\bT} { \mbox{\boldmath $T$}}
\newcommand{\bU} { \mbox{\boldmath $U$}}
\newcommand{\bV} { \mbox{\boldmath $V$}}
\newcommand{\bW} { \mbox{\boldmath $W$}}
\newcommand{\bX} { \mbox{\boldmath $X$}}
\newcommand{\bY} { \mbox{\boldmath $Y$}}
\newcommand{\bZ} { \mbox{\boldmath $Z$}}
\newcommand{\bzero} { \mbox{\boldmath $0$}}
\newcommand{\bfell} {\mbox{\boldmath $ \ell $}}

%
% special math symbols
%
\newcommand{\dou}{\partial}
\newcommand{\leftjb} {[\![}
\newcommand{\rightjb} {]\!]}
\newcommand{\ju}[1]{ \leftjb #1 \rightjb }
\newcommand{\D}[1]{\mbox{d}{#1}} 
\newcommand{\grad}{\mbox{\boldmath $\nabla$}}
\newcommand{\modulus}[1]{|#1|}
\renewcommand{\div}[1]{\grad \cdot #1}
\newcommand{\curl}[1]{\grad \times #1}
\newcommand{\mean}[1]{\langle #1 \rangle}
\newcommand{\bra}[1]{{\langle #1 |}}
\newcommand{\ket}[1]{| #1 \rangle}
\newcommand{\braket}[2]{\langle #1 | #2 \rangle}
\newcommand{\dbdou}[2]{\frac{\dou #1}{\dou #2}}
\newcommand{\dbdsq}[2]{\frac{\dou^2 #1}{\dou #2^2}}
\newcommand{\Pint}[2]{ P \!\!\!\!\!\!\!\int_{#1}^{#2}}
\newcommand{\Itwo}{{\mathds{1}}}
\newcommand{\Hds}{{\mathds{H}}}
\newcommand{\cH}{{\cal H}}
\newcommand{\cS}{{\cal S}}

%
% Abbreviations for equations etc
% 
\newcommand{\prn}[1] {(\ref{#1})}
\newcommand{\sect}[1] {Section~\ref{#1}}
\newcommand{\Sect}[1] {Section~\ref{#1}}

%
% Other utilities
%
\newcommand{\uncon}[1]{\centerline{\epsfysize=#1 \epsfbox{/usr2/yogeshwar/styles/construction.pdf}}}
\newcommand{\checkup}[1]{{(\tt #1)}\typeout{#1}}
\newcommand{\ttd}[1]{{\color[rgb]{1,0,0}{\bf #1}}}
\newcommand{\red}[1]{{\color[rgb]{1,0,0}{\protect{#1}}}}
\newcommand{\blue}[1]{{\color[rgb]{0,0,1}{#1}}}
\newcommand{\green}[1]{{\color[rgb]{0.0,0.5,0.0}{#1}}}
\newcommand{\citebyname}[1]{\citeauthor{#1}\cite{#1}}
\newcommand{\myfigwidth}{0.95\columnwidth}
\newcommand{\myhalffig}{0.475\columnwidth}
\newcommand{\mythirdfig}{0.33\columnwidth}
\newcommand{\signum}[0]{\mathop{\mathrm{sign}}}
\newcommand{\skup}{\ket{s \uparrow}}
\newcommand{\skdn}{\ket{s \downarrow}}
\newcommand{\pkup}{\ket{p \uparrow}}
\newcommand{\pkdn}{\ket{p \downarrow}}
\newcommand{\sbup}{\bra{s \uparrow}}
\newcommand{\sbdn}{\bra{s \downarrow}}
\newcommand{\pbup}{\bra{p \uparrow}}
\newcommand{\pbdn}{\bra{p \downarrow}}

% Abbreviations for equations etc
\newcommand{\Eqn}[1] {Eqn.~(\ref{#1})}
\newcommand{\Fig}[1]{Fig.~\ref{#1}}

%\title{Exact relations between quantum coherence and measure of localization: application to many-body localization transition}
\title{Trade-off relations between quantum coherence and measure of many-body localization}
\author{Arti Garg}
\affiliation{Theory Division, Saha Institute of Nuclear Physics, 1/AF Bidhannagar, Kolkata 700 064, India}
\affiliation{Homi Bhabha National Institute, Training School Complex, Anushaktinagar, Mumbai 400094, India}
\author{Arun Kumar Pati}
\affiliation{Centre for Quantum Engineering, Research and Education (CQuERE), TCG CREST, Kolkata, India}
\vspace{0.2cm}
\begin{abstract}
\vspace{0.3cm}
{Quantum coherence, a fundamental resource in quantum computing and quantum information, often competes with localization effects that affects quantum states in disordered systems.
%A measure of localization and many-body localization (MBL), such as the inverse participation ratio, is crucial for understanding the complex interplay between quantum information and condensed matter physics. 
In this work, we prove exact trade-off relations between quantum coherence and a measure of localization and many-body localization, namely, the inverse participation ratio (IPR). We prove that for a pure quantum state, $l_1$-norm of quantum coherence and the relative entropy of coherence satisfy complementarity relations with IPR.
For a mixed state, IPR and the $l_2$-norm of  quantum coherence as well as relative entropy of coherence satisfy trade-off inequalities. %These relations suggest that quantum coherence in disordered quantum systems is also an ideal characterization of the delocalisation to many-body localisation transition, much like IPR, which is a well-known diagnostic of MBL.
 We further explore the applicability of these general relations between quantum coherence and measure of localization in the context of delocalization to many-body localization transition in a disordered interacting quantum many-body system and demonstrate that quantum coherence is also an ideal characterization of the delocalisation to many-body localisation transition. Our numerical analysis shows that though the full system loses coherence in the MBL phase, coherence of a subsystem can increase with the combined effect of disorder and interactions. %Additionally, delocalization to MBL transition point obtained from the relative entropy of coherence and $l_2$ norm of coherence for a subsystem is consistent with the transition point obtained from conventional diagnostics of MBL like level-spacing ratio and bipartite entanglement entropy. 
 %These relations also provide insight into the unusual properties of bipartite entanglement entropy across the MBL transition. 
 We believe that these trade-off relations can help in better understanding of how coherence can be preserved or lost in realistic many-body quantum systems, which is vital for developing robust quantum technologies and uncovering new phases of quantum matter.}
        
\end{abstract} 
\maketitle
\section{Introduction}
Many-body localisation (MBL) is a fascinating phenomenon at the intersection of quantum condensed matter physics, quantum statistical physics, and quantum information science~\cite{Huse_rev,Alet_rev,Abanin_rev,Vidmar_rev}.
A system of interacting particles in the MBL phase fails to reach thermal equilibrium due to the presence of strong disorder, effectively ``freezing'' the system in a non-ergodic state which also results in long memory of initial states~\cite{expt1,expt3,Soumya,Titas,Gornyi,yp}. A system in the MBL phase exhibits unusual features of entanglement entropy(EE)~\cite{Abanin_rev}; even highly excited eigenstates of a system in the MBL phase have area law of EE~\cite{Huse,Sdsarma,Nayak,garg}. Starting from a product initial state, a unitary time evolution with respect to the Hamiltonian of an MBL system leads to the slow growth of the bipartite EE with time~\cite{Moore,Serbyn_EE,garg_lr,yp,expt_EE}. %Entanglement entropy is therefore one of the most interesting characterization of the MBL phase. 
Can quantum coherence provide a characterization of the delocalization to MBL transition? Since entanglement is a way to express a physical system's quantum-ness and is known to originate from the superposition principle—which is also a necessary component of coherence—it is an intriguing question to ask. Our study addresses this query by establishing exact relations between the measure of MBL and quantum coherence, as well as revealing various norms of quantum coherence as a novel characterization of the delocalization to MBL transition.

Quantum coherence is a fundamental notion in quantum mechanics. It is a crucial resource for various quantum technologies, including quantum computing, quantum communication, and quantum sensing. It represents the ability of a quantum system to exhibit superposition, which underpins the advantage that quantum systems have over classical counterparts. It is frequently a precondition for entanglement and other types of quantum correlations. However, in real-world systems, especially those involving many interacting particles, maintaining quantum coherence becomes a significant challenge due to the presence of disorder and interactions that can
lead to many-body localization. As we would show later, though MBL can protect certain quantum states from decoherence, it also poses a threat to quantum coherence by confining quantum states to localized regions of the system, thus preventing the spread of quantum information. To be specific, we demonstrate that the coherence of a subsystem can be maintained during the MBL phase, despite the loss of coherence for the entire system.

  A rigorous framework for characterising quantum coherence has been  proposed recently~\cite{coh,coh_Shao,Cheng}. Quantum coherence, like IPR, is a basis-dependent quantity. In the past, relationships between quantum coherence measurements in different reference bases have been explored ~\cite{Pati_2016}. This is important for connecting information across distinct basis states. The amount of coherence that a quantum system can possess is limited by its mixedness which itself depends upon the environmental noise~\cite{Pati_mixedness}. Coherence can be estimated by non-commutativity of any observable with its ``incoherent-part''~\cite{pati_2022} and is directly related to the EE~\cite{coh_ee1} as well as 'magic' of a quantum state~\cite{Pati_2018}. Recent studies have shown that entanglement in a system can be utilised to assess quantum coherence ~\cite{coh_ee2,coh_ee3}. The correlation between coherence and entanglement suggests a direct relationship between localisation and coherence that we explore in this work. %We investigate quantum coherence in MBL systems to answer two significant questions: how quantum coherence describes delocalisation to MBL transition and how MBL preserves quantum state coherence.

In this work, we use inverse participation ratio (IPR) as the measure of localization and derive following trade-off relations between quantum coherence and IPR.Though both IPR and coherence are basis dependent quantities, the exact relations established here are basis independent.
For any pure state following relations hold: (i) $ C_1 + IPR \ge 1$ where $C_1$ is the $l_1$ norm of the quantum coherence and (ii) $1 \le C_{rel}+IPR \le d$ where $C_{rel}$ is the relative entropy of coherence and $d$ is the dimension of the Hilbert space. For a mixed state with density matrix $\rho$ various measures of coherence satisfy trade-off relations with IPR: (i) $C_2(\rho) +IPR(\rho) \le 1$ where $C_2(\rho)$ is the $l_2$ norm of the quantum coherence, (ii) $C_{rel}(\rho)+IPR(\rho)+S(\rho) \ge 1$ and (iii) $C_{rel}(\rho)+d_n~IPR(\rho) \ge 1$ where $C_{rel}(\rho)$ is the relative entropy of coherence, $S(\rho)$ is  the von Neumann entropy  and $d_n$ is the number of non-zero eigenvalues of $\rho$. We also found an upper bound on relative entropy of coherence for the mixed state which is given by $C_{rel}(\rho)+IPR(\rho)+M(\rho)\le d$ where $M(\rho)=1-Tr(\rho^2)$ is the mixedness of the state and $d$ is the dimension of $\rho$. To the best of our knowledge these relations have not been reported earlier.% In the sequel, we provide details about these inequalities and analyse them in the context of MBL transition. 

Understanding the trade-off between quantum coherence and localization is critical for several reasons. Firstly, in the development of quantum technologies, it is important to know how disorder and interactions can either preserve or degrade quantum coherence. If localization effects can be harnessed to protect coherence in certain scenarios, this could lead to new methods for preserving quantum information in noisy or disordered environments. On the other hand, if localization excessively restricts coherence, it may necessitate strategies to mitigate these effects, such as through error correction or system engineering. Secondly, the study of these trade-off relations contributes to our fundamental understanding of the transition from ergodic to non-ergodic phases in many-body systems. The interplay between coherence and localization may reveal new quantum phases or critical points that are not accessible through classical means. 
 Furthermore, exploring these trade-offs can provide insights into the stability of quantum coherence in various quantum materials, which is essential for the design of new quantum devices.
 
The rest of the manuscript is organized as follows. In section 2 we prove exact relations between various norms of coherence and IPR for pure and mixed states. These relations are generic and apply to any quantum system. In the next section we analyze these relations for a model of spinless fermions with random disorder and interactions followed up by the analysis of various norms of coherence across the delocalization to MBL transition. At the end we conclude by emphasizing the relevance of relations between MBL and coherence in the context of quantum devices.

\section{ General relations between coherence and measure of localization}
In this section we derive generic trade-off relations between various measures of quantum coherence and measure of localization, which hold true for generic quantum states.
Let us recapitulate some basic definitions.  We use various measures of quantum coherence, which are defined below. $l_1$ norm of quantum coherence is defined as $C_{1}=\sum_{\alpha \ne \beta}|\rho_{\alpha\beta}|$ where $\rho$ is the density matrix for the quantum state under consideration. Relative entropy of coherence, which satisfies a trade-off relation with disturbance caused by measurement~\cite{Pati_PRA2018} and provides an operational coherence measure~\cite{Pati_PRL} is defined as $C_{rel}=-\sum_{\alpha}\rho_{\alpha\alpha}\log\rho_{\alpha\alpha}-S(\rho)$ with $S(\rho)=-Tr[\rho \log\rho]$ is the von Neumann entropy~\cite{coh}.  Here $\log$ is w.r.t base 2. Both these measures of coherence fulfill the following criteria: (1) $C(\rho)=0$ for incoherent states, that is, states that have a diagonal density matrix in a fixed basis otherwise $C(\rho) > 0$. (2) Under selective incoherent operations $C(\rho)$ does not increase. (3) $C(\rho)$ is a convex function of quantum states, that is, $C(\sum_k p_k \rho_k) \le \sum_k p_k C(\rho_k)$. We also use  $l_2$ norm of quantum coherence which is defined as $C_{2}=\sum_{\alpha \ne \beta}|\rho_{\alpha\beta}|^2$ and has been shown to be equal to the infinite-time averaged return probability~\cite{Namit2019}. Note that though $l_2$ norm of coherence does not satisfy monotonicity~\cite{coh}, it still provides an ideal probe to study many-body localization as we discuss later in this work. We use inverse participation ratio (IPR) as the measure of localization. IPR is a widely used measure to quantify the degree of localization, with higher IPR values indicating stronger localization.\\

{ \it Relations between measure of localization and quantum coherence for a pure state:} Let us consider an isolated many-body quantum system with a finite dimensional Hilbert space of dimension $N_F$.
Consider a pure quantum state $|\Psi\ra \in {\cal H}^N_F$ with $|\Psi\ra = \sum_\alpha a_\alpha |\alpha\ra$, where $\{ |\alpha\ra \}$ are the basis states. 
The corresponding density matrix is $\rho=|\Psi\ra \la \Psi|$ such that $\rho_{\alpha\beta} =a_{\alpha}a_\beta^\star$.
IPR for this state is defined as
\be
IPR(\Psi)=\sum_{\alpha} |\la \alpha |\Psi \ra|^4.
\label{ipr}
\ee
Let us consider the relative entropy of coherence for a pure state. Since $S(\rho)=0$ for a pure state,
        \be
        C_{rel}(\Psi) 
        %=-\sum_{\alpha}\rho_{\alpha,\alpha}\log \rho_{\alpha,\alpha} 
        = -\sum_\alpha|a_\alpha|^2\log |a_\alpha|^2.
        \label{coh_rel}
        \ee
         For any positive $x$, $-\log (x) \ge 1-x$. Using $x = |a_\alpha|^2$ in this logarithmic inequality, we have
         \begin{align}
          C_{rel}(\Psi) \ge \sum_{\alpha}|a_\alpha|^2(1-|a_\alpha|^2)= 1-IPR(\Psi) \nonumber \\
         C_{rel}(\Psi)+IPR(\Psi) \ge 1.
        \label{reln2}
        \end{align}
 The  $l_1$ norm of coherence for this pure state is defined as 
\be
C_1(\Psi) =\sum_{\alpha \ne \beta}|a_\alpha||a_\beta|.
\label{C1}
\ee\href{}{}

For a normalized state $|\Psi\ra$, $\sum_{\alpha}|a_\alpha|^2\sum_\beta|a_\beta|^2 = \sum_\alpha|a_\alpha|^4+\sum_{\alpha \ne \beta} |a_\alpha|^2|a_\beta|^2=1$.
        Since $|a_\alpha|^2$ is the probability of being in the state $\alpha$ of the Fock space, we have  $|a_\alpha|^2 \le |a_\alpha|~~\forall \alpha$. Using this we obtain the another trade-off relation between IPR and coherence as given below: 
        \begin{align}
        IPR(\Psi) +C_1(\Psi) \ge 1.
        \label{reln1}
        \end{align}

\indent{\it Relation between IPR and quantum coherence for a mixed state}:  Given a many-body system, if we are interested in any subsystem, then that can be described a mixed state density matrix $\rho$. The inverse participation ratio (IPR) for a density operator $\rho$  is defined as 
\be
IPR(\rho)=\sum_\alpha |\rho_{\alpha\alpha}|^2.
\label{ipr_rho}
\ee 
Since $~Tr (\rho)=1$, we have
\begin{align}
1 =\sum_\alpha |\rho_{\alpha\alpha}|\sum_\beta |\rho_{\beta\beta}|=\sum_{\alpha}|\rho_{\alpha\alpha}|^2+\sum_{\alpha \ne \beta}|\rho_{\alpha\alpha}| |\rho_{\beta\beta}|.
\label{rho_tr}
\end{align}
For any positive-definite matrix $\rho$, we have 
\be
\sqrt{\rho_{\alpha\alpha}}\sqrt{\rho_{\beta\beta}} \ge |\rho_{\alpha\beta}| .
%\imples |\rho_{\alpha\alpha}| |\rho_{\beta\beta}| \ge |\rho_{\alpha\beta}|^2.
\label{rho2}
\ee

Using condition ~(\ref{rho2}) in Eq.~(\ref{rho_tr}), we obtain 
\begin{align}
1 & \ge \sum_\alpha |\rho_{\alpha\alpha}|^2 + \sum_{\alpha \ne\beta}|\rho_{\alpha\beta}|^2 = IPR(\rho)+C_2(\rho) \le 1,
\label{reln3}
\end{align}
where $C_2(\rho)=\sum_{\alpha \ne \beta} |\rho_{\alpha\beta}|^2$ is the $l_2$ norm of quantum coherence. This relation provides an upper bound on the value of the $l_2$ norm of coherence for a mixed state. It is easy to see that equality in relation ~(\ref{reln3}) holds true for a pure state~\cite{Namit2021}.
\begin{figure*}
  \begin{center}
  \includegraphics[width=3.0in,angle=0]{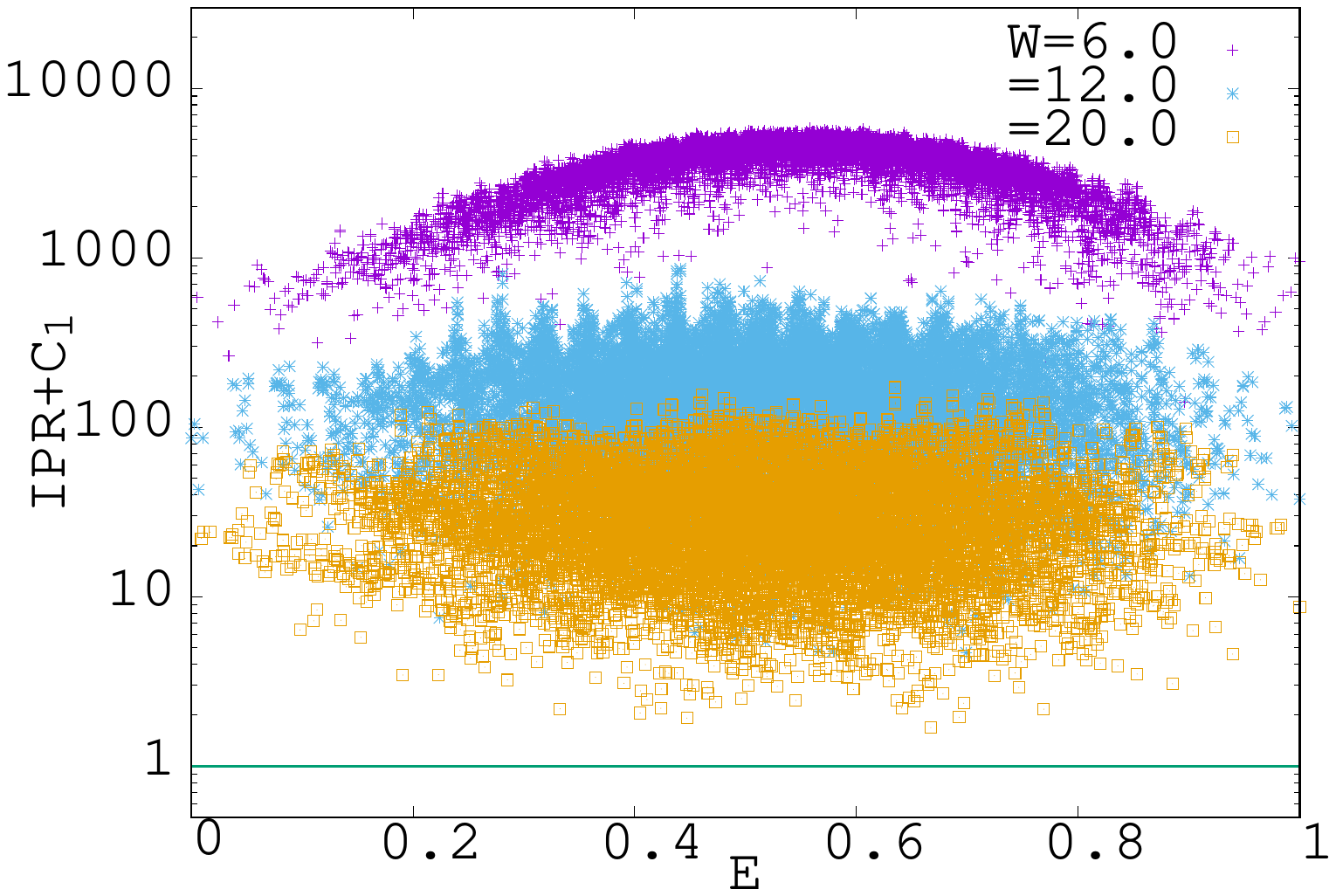}
  \includegraphics[width=3.0in,angle=0]{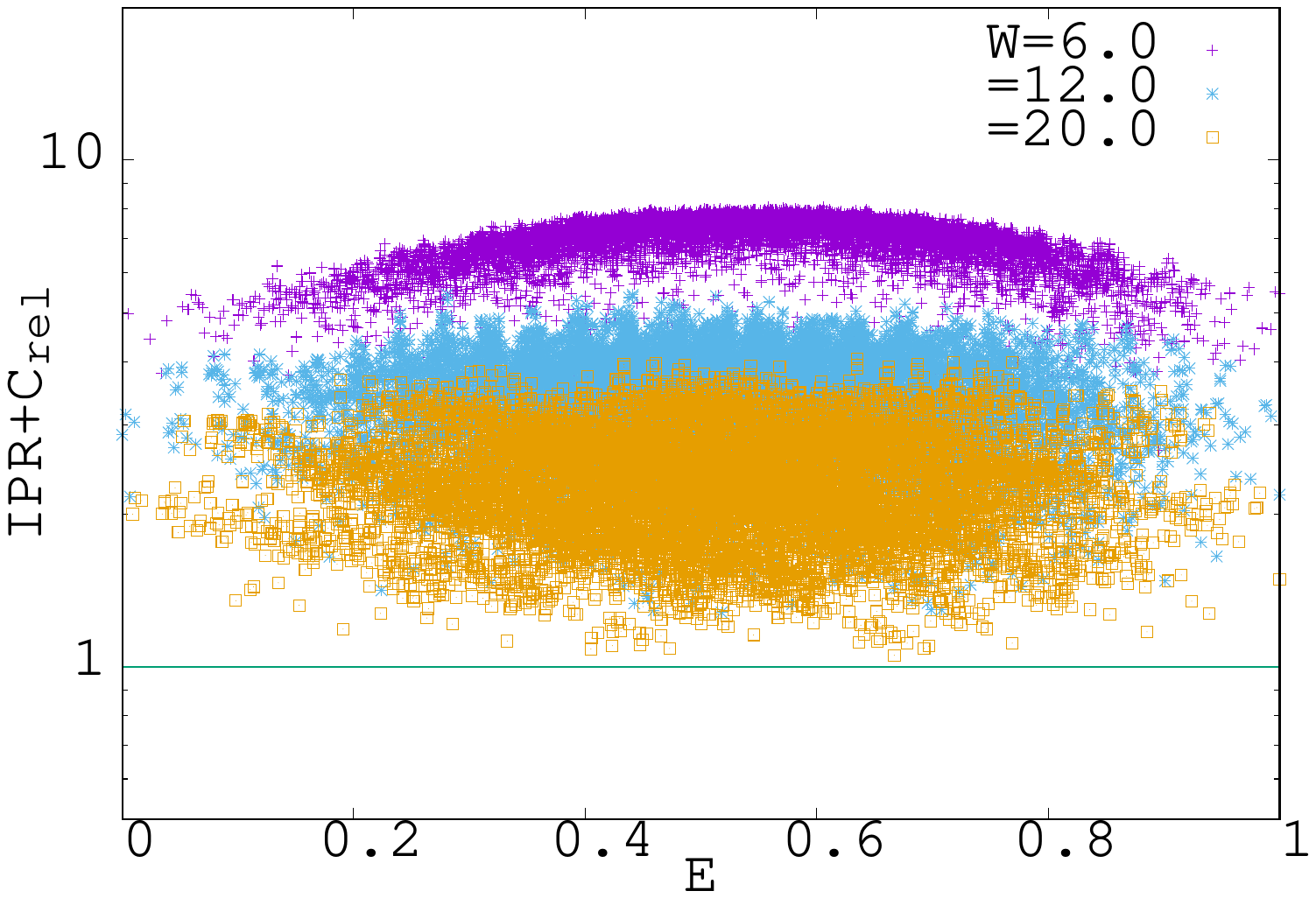}
  \caption{Left panel shows the plot of $IPR +C_1$ as a function of energy $E$ for various disorder values. The right panel shows $IPR+C_{rel}$ calculated for the whole system for various disorder values. As one can see, $IPR+C_1$ is always greater than one for any eigenstate and for any disorder configuration. As $W$ increases, the sum approaches unity. The same holds true for $IPR+C_{rel}$. The data shown is for one disorder configuration.}
  \label{pure} 
  \end{center}
   \end{figure*}
Now we prove a trade-off relation between the relative entropy of coherence $C_{rel}(\rho)$ and IPR for a mixed state. For a mixed state $\rho$, we have 
\bea
C_{rel}(\rho)= -\sum_{\alpha}\rho_{\alpha \alpha}\log \rho_{\alpha \alpha} +Tr [\rho \log \rho].
\eea
Let $\{\lambda_n\}$ represent the eigenvalues of the density matrix $\rho$. Then von Neumann entropy is $S(\rho)=-\sum_n \lambda_n \log(\lambda_n)$. Consider, the logarithmic inequality 
\be
\frac{b-a}{a}-\log{b}+\log{a}\le \frac{(b-a)^2}{ab},
\ee
which holds true for any $a,b >0$.  On Substituting $a=\rho_{\alpha\alpha}$ and $b=\lambda_n$, one gets,
\bea
-\rho_{\alpha\alpha}[\lambda_n \log{\lambda_n}]+\lambda_n [\rho_{\alpha \alpha}\log{\rho_{\alpha\alpha}}]\le \rho_{\alpha \alpha}^2-\lambda_n\rho_{\alpha \alpha} .
\eea
Summing over the index $\alpha$ in the above inequality  and using $\sum_\alpha \rho_{\alpha\alpha} =1$, gives us a relation 
\be
-(\lambda_n \log{\lambda_n})-\lambda_n S(\rho_D) \le IPR(\rho)-\lambda_n.
\ee
Here $S(\rho_D)=-\sum_{\alpha}\rho_{\alpha\alpha}\log \rho_{\alpha\alpha}$ is the Shannon entropy. Now summing over the eigenvalue index $n$, we get
\begin{align}
S(\rho)-S(\rho_D) \le d_n *IPR(\rho)-1 \nonumber \\ 
 C_{rel}(\rho)+ d_n *IPR(\rho) \ge 1.
\label{reln4}
\end{align}

 Here, $d_n$ is the number of non-zero eigenvalues of $\rho$ and $\sum_n \lambda_n =1$. Note that the relation~[\ref{reln4}] maps onto the relation~[\ref{reln2}] for a pure state for which $d_n =1$. 

Another interesting relation between $C_{rel}$ and IPR can be obtained by using the identity $-\log(\rho_{\alpha \alpha}) \ge 1-\rho_{\alpha \alpha}$ in the definition of $C_{rel}$ as follows:
\begin{align}
C_{rel}(\rho) +S(\rho) =-\sum_{\alpha}\rho_{\alpha\alpha}\log \rho_{\alpha\alpha} \ge \sum_{\alpha} \rho_{\alpha \alpha}(1-\rho_{\alpha \alpha})  \nonumber \\
  C_{rel}(\rho) + S(\rho) +IPR(\rho)  \ge 1.
\label{reln5}
\end{align}

One can also determine an upper bound on $C_{rel}(\rho)$. Using the identity $\log (\rho) \le \rho-\mathbb{I}$,
we have 
\begin{align}
C_{rel}(\rho)  \le -\sum_{\alpha} \rho_{\alpha \alpha}\log\rho_{\alpha \alpha}+Tr \rho(\rho-\mathbb{I}) \nonumber \\
 C_{rel}(\rho) \le \eta(\rho)-1-\sum_{\alpha}\rho_{\alpha \alpha}\log \rho_{\alpha \alpha},
\end{align}
where $\eta(\rho)=Tr \rho^2$ is the purity of the mixed state $\rho$. Since $x \log{x} \ge x-1$ for all $x>0$, and for the 
case of $x <1$, we have $x  \ge x^2$ and one can use the inequality $x \log{x} \ge x^2-1$. 

Now choosing $x = \rho_{\alpha\alpha}$ in the above identity implies that 
\be
-\sum_{\alpha} \rho_{\alpha\alpha} \log \rho_{\alpha\alpha} \le d-IPR(\rho).
\label{log_sd}
\ee
Here $d$ is the dimension of the density matrix $\rho$. 
%Since $log_2\rho_{\alpha \alpha} \le \rho_{\alpha \alpha}-1$ which holds because $\rho_{\alpha \alpha}>0$, one gets
Using this bound on the Shannon entropy in Eq.~(\ref{log_sd}), we obtain
\bea
C_{rel}(\rho)+ IPR(\rho) + M(\rho) \le d,
\label{reln6}
\eea
where $M(\rho) =(1- \eta(\rho) )$ is the mixedness. This can be interpreted as a tradeoff relation between coherence, IPR and mixedness. Also, this proves an upper bound for the sum of the quantum coherence and IPR. 
Here $d$ is the dimension of the density matrix $\rho$.  For a pure state, since $\eta(\rho)=1$, $C_{rel}(\rho)+IPR(\rho) \le d$.

\begin{figure}[]
  \begin{center}
%    \hspace{-1cm}
  \includegraphics[width=3.2in]{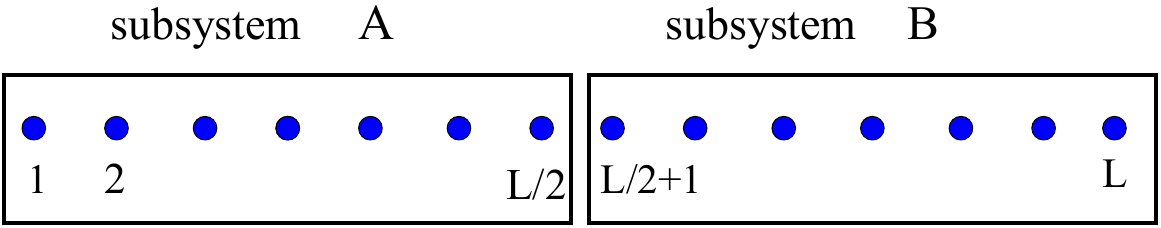}
   \caption{An illustration of the two subsystems used to calculate the sub-lattice quantum coherence and the bipartite entanglement entropy.}
      \label{EE}
\vskip-1cm
\end{center}
\end{figure}
            
\section{Tradeoff relations between coherence and localization in a disordered interacting system}
In this section, we calculate various measures of coherence for a disordered interacting system and look for situations where the above derived inequalities provide strict bounds on coherence.  %and compare the results with standard diagnostics of MBL to %demonstrate the usefulness  of quantum coherence in %characterising the MBL transition.  

To be specific, we study quantum coherence in the standard model of many-body localization, namely, a model of spin-less fermions in one-dimension described by the following Hamiltonian  
\bea
H=-t\sum_{i}[c^\dagger_ic_{i+1}+h.c.] + \sum_i h_i n_i \nonumber \\
+\sum_{i} V n_in_{i+1} +V_2 n_i n_{i+2}
\label{model}
\eea
with periodic boundary conditions. Here $t$ is the nearest neighbor hopping amplitude, $V$ is the strength of nearest neighbour repulsion between Fermions and $V_2$ is the strength of next-nearest-neighbour repulsion among fermions.  The onsite potential $h_i \in [-W/2,W/2]$ is uniformly distributed with $W$ as the disorder strength. We study this model at half-filling of fermions using exact diagonalization. In the entire analysis we fix $V=t(=1)$ and $V_2=0.5t$. Note that a similar model of spins-1/2 was studied earlier~\cite{Prosen} where both the spin-flip as well as spin-preserving terms were included up to second neighbour. But in the model we are studying, only the interaction between spinless fermions is up to second neighbour while the hopping of fermions is limited to first neighbours only. 
 \begin{figure*}
  \begin{center}
  \includegraphics[width=3.0in,angle=0]{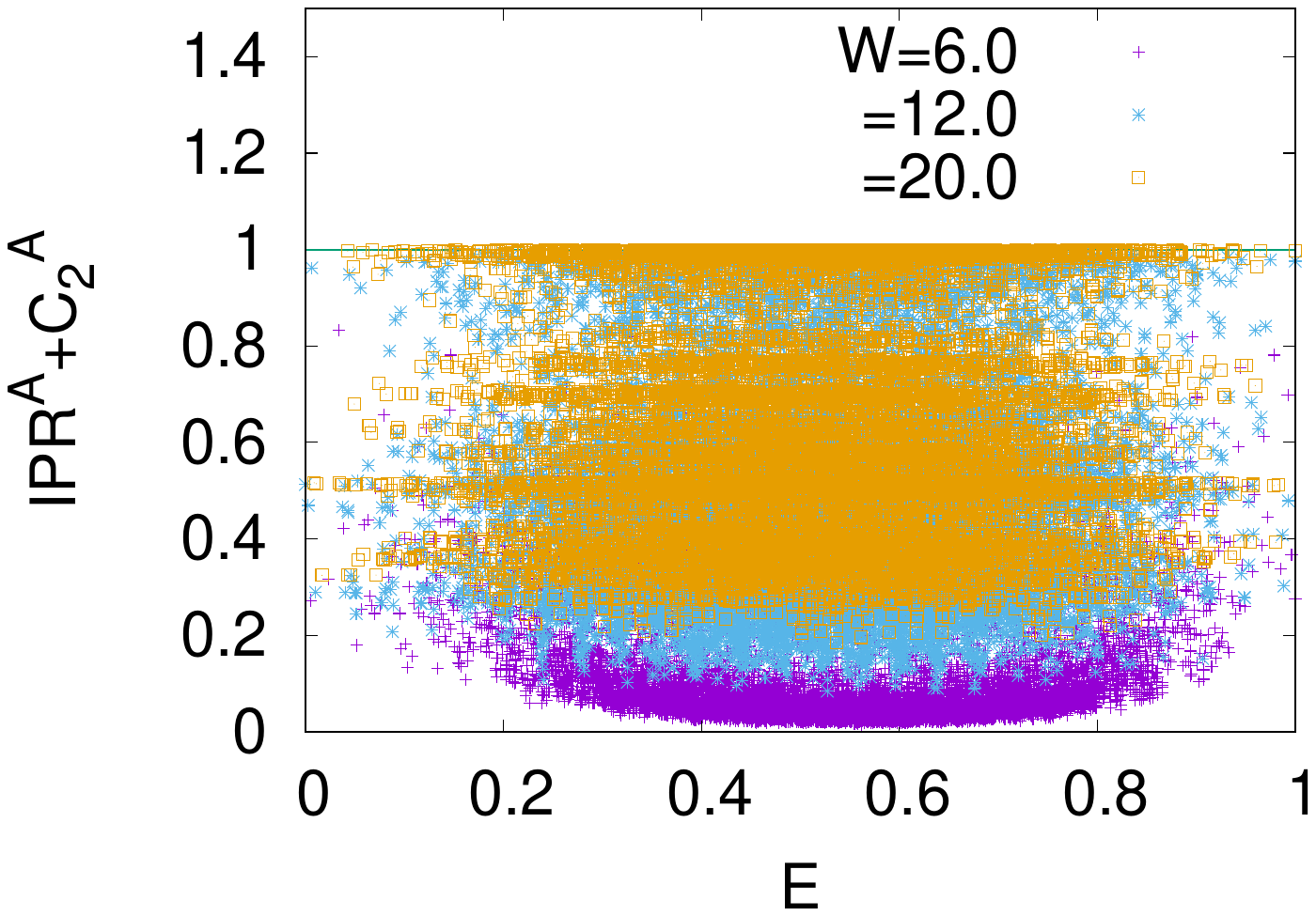}
  \includegraphics[width=3.0in,angle=0]{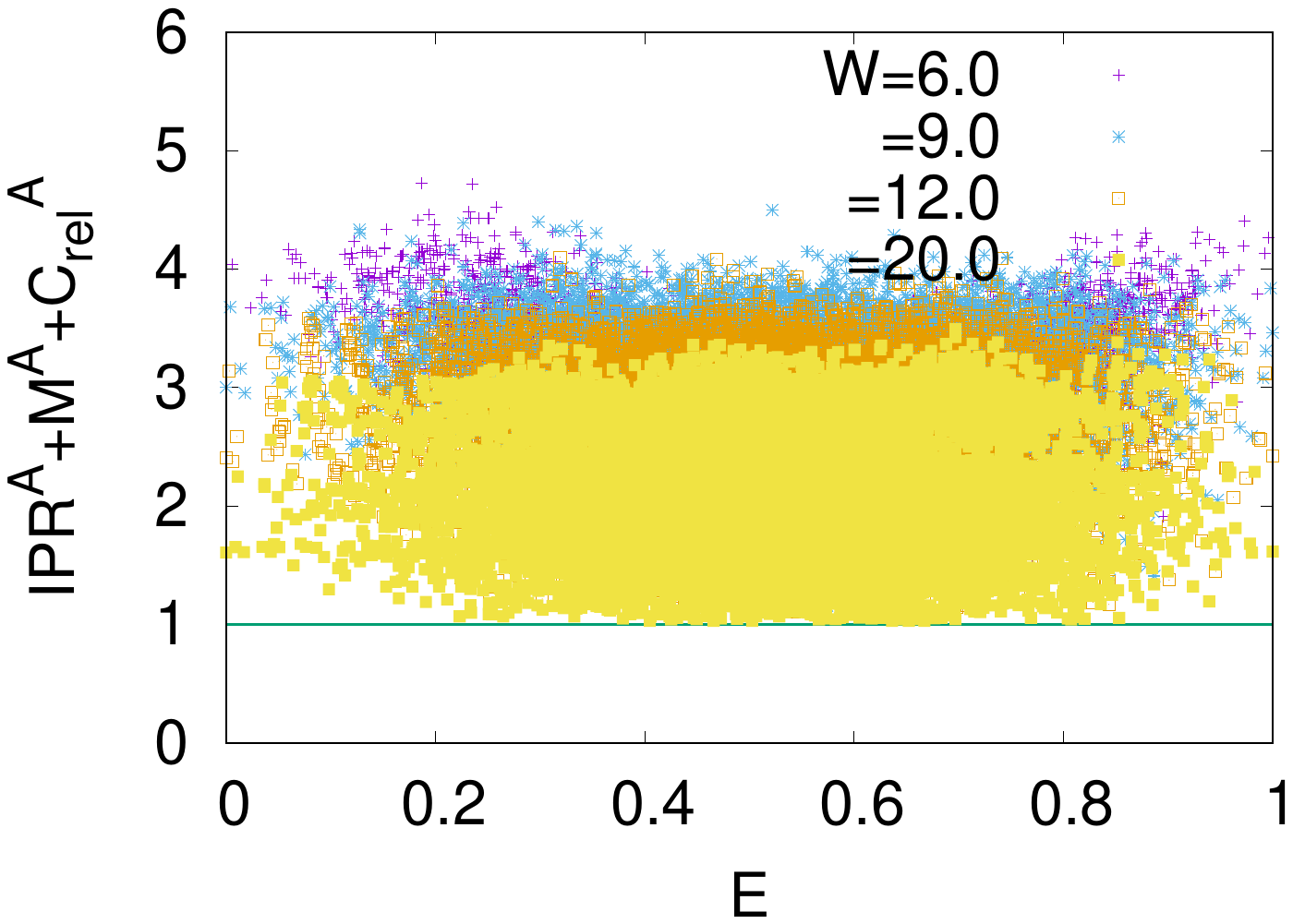}
  \caption{The left panel shows $IPR^A +C_2^A$ vs rescaled energy $E$ for various disorder values. $IPR^A+C_2^A$ is less then one for any eigenstate for any disorder configuration and approaches one as $W$ increases. On the other hand $IPR^A+M+C_{rel}^A$ is bounded from above by $d$, dimension of the reduced density matrix. As $W$ increases the sum $IPR^A+M+C_{rel}^A$ approaches unity.}
  \label{mix}
  \end{center}
   \end{figure*}

In the weak disorder limit of the model in Eq.~(\ref{model}), most of the many-body eigenstates are extended. For a conventionally extended state, $|\Psi\ra = \frac{1}{\sqrt{N_F}}\sum_{\alpha=1}^{N_F}|\alpha\ra$,  $IPR =\frac{1}{N_F}$ which goes to zero in the $N_F\rightarrow \infty$ limit, where $N_F$ is the dimension of the Hilbert space. In this limit relation ~[\ref{reln1}] implies that $C_1$ can not be zero for an extended state. In fact for a conventionally extended state $l_1$ norm of quantum coherence $C_1 =N_f-1 \gg 1$. In Fig.~\ref{pure} we have shown $IPR+C_1$ vs eigenenergy $E =\frac{E_n-E_{min}}{E_{max}-E_{min}}$ for various disorder values. For $W=6.0t$, $IPR+C_1$ is much larger than one but as the disorder strength increases $IPR+C_1$ approaches one for most of the eigenstates of the system under consideration. This holds true for any random disorder configuration. Since for a localized state, which has contributions from only a small fraction $N_{occ}$ out of $N_F$ states in the Hilbert space, $IPR \sim 1/N_{occ}$ and $C_1 \sim N_{occ}-1$ such that for an extremely localized state, $IPR +C_1$ approach unity as shown in Fig,~\ref{pure}. 

A similar trend is seen for the relative entropy of coherence. Relative entropy of coherence for a maximally extended state is $C_{rel} = -\sum_\alpha \frac{1}{N_F}\log\frac{1}{N_F} = log N_F$. Therefore, $C_{rel}+IPR= \log N_F +1/N_F \gg 1$ for states which are extended as shown in the right panel of Fig.~\ref{pure}. As $W$ increases, $IPR+C_{rel}$ decreases and approaches unity for all the eigenstates in the limit of very large disorder for any disorder configuration. For a localized state $C_{rel}=\log (N_{occ})$ and in the limit of extremely localized states $C_{rel} +IPR \rightarrow 1$ with $C_{rel}$ tending to zero.         

 To explore the effect of disorder on coherence for mixed states, we divided the system into two subsystems $A$ and $B$ as shown in Fig.~[\ref{EE}]. For any eigenstate $|\Psi_n\ra$ of the Hamiltonian in Eq.~(\ref{model}), the corresponding reduced density-matrix for subsystem $A$ is defined as $\rho^A = Tr_B [|\Psi_n\ra\la \Psi_n|]$. $\rho^A$ represents a mixed state and should obey relations ~(\ref{reln3},\ref{reln4},\ref{reln5} and \ref{reln6})  between the IPR of the sublattice $A$ and various measures of coherence for sublattice $A$. Since bipartite entanglement entropy of an MBL system has unusual properties, it would be interesting to use above mentioned inequalities to explore quantum coherence for a bipartite system.

The left panel of Fig.~\ref{mix} shows $IPR^A+C_2^A$ as a function of eigenenergy $E$ for various values of $W$ but for any random disorder configuration. For weak disorder, $IPR^A+C_2^A$ is close to zero and as disorder strength increases it approaches unity. To understand this trend, again consider a maximally extended eigenstate as above. For a system of $L$ sites the reduced density-matrix for the subsystem $A$ is a $2^{L/2} \times 2^{L/2}$ matrix with $\rho_{\alpha \beta}^A \propto 1/N_F$, $\forall \{\alpha,\beta\}$. Thus, IPR for the subsystem $A$ is $IPR^A \sim \lbr \frac{1}{N_F}\rbr^2 2^{L/2}$ which goes to zero for $N_F \rightarrow \infty$. The $l_2$ measure of quantum coherence for the subsystem $A$ is $C_2^{A}  \sim \lbr \frac{1}{N_F}\rbr^2 (2^{L/2}*(2^{L/2}-1))$ which also goes to zero as $N_F$ increases making $IPR^A+C_2^A$ vanishingly small. Thus, for a maximally extended state sublattice coherence goes to zero in the infinite system size limit which is in complete contrast to the coherence of the full system. 
For a localized eigenstate $|\Psi\ra = \frac{1}{\sqrt{N_{occ}}} \sum_{\alpha=1}^{N_{occ}}|\alpha \ra$ with $N_{occ} \ll N_F$, reduced density-matrix for subsystem $A$ is $\rho_{\alpha\beta}^A \sim \frac{1}{N_{occ}}$ such that $IPR^A +C_2^{A}$ approach unity in the strong disorder limit with both $IPR^A$ and $C_2^A$ increasing with increase in disorder and decrease in $N_{occ}$. 

The right  panel of Fig.~\ref{mix} shows $IPR^A+M^A+C_{rel}^A$ for various values of $W$ for a random disorder configuration. Here $M$ is mixedness of the reduced density matrix $\rho^A$. For weak disorder, $IPR^A+M^A+C_{rel}^A$ is much greater than one being consistent with the relation~(\ref{reln6}). As $W$ increases, this sum decreases and approaches unity in the limit of very strong disorder. Further, since $IPR^A+M^A$ decreases (which happens mainly due to suppression in mixedness) as $W$ increases, $C_{rel}^A$ increases with increase in $W$.

%Thus, all the relations proved in earlier section for the pure as well as mixed states, approach unity for extremely localized states. 

\section{Quantum coherence across the MBL transition}
\begin{figure}[ht]
  \begin{center}
    \vskip 0.5cm
    \hspace{-1cm}
  \includegraphics[width=3.5in]{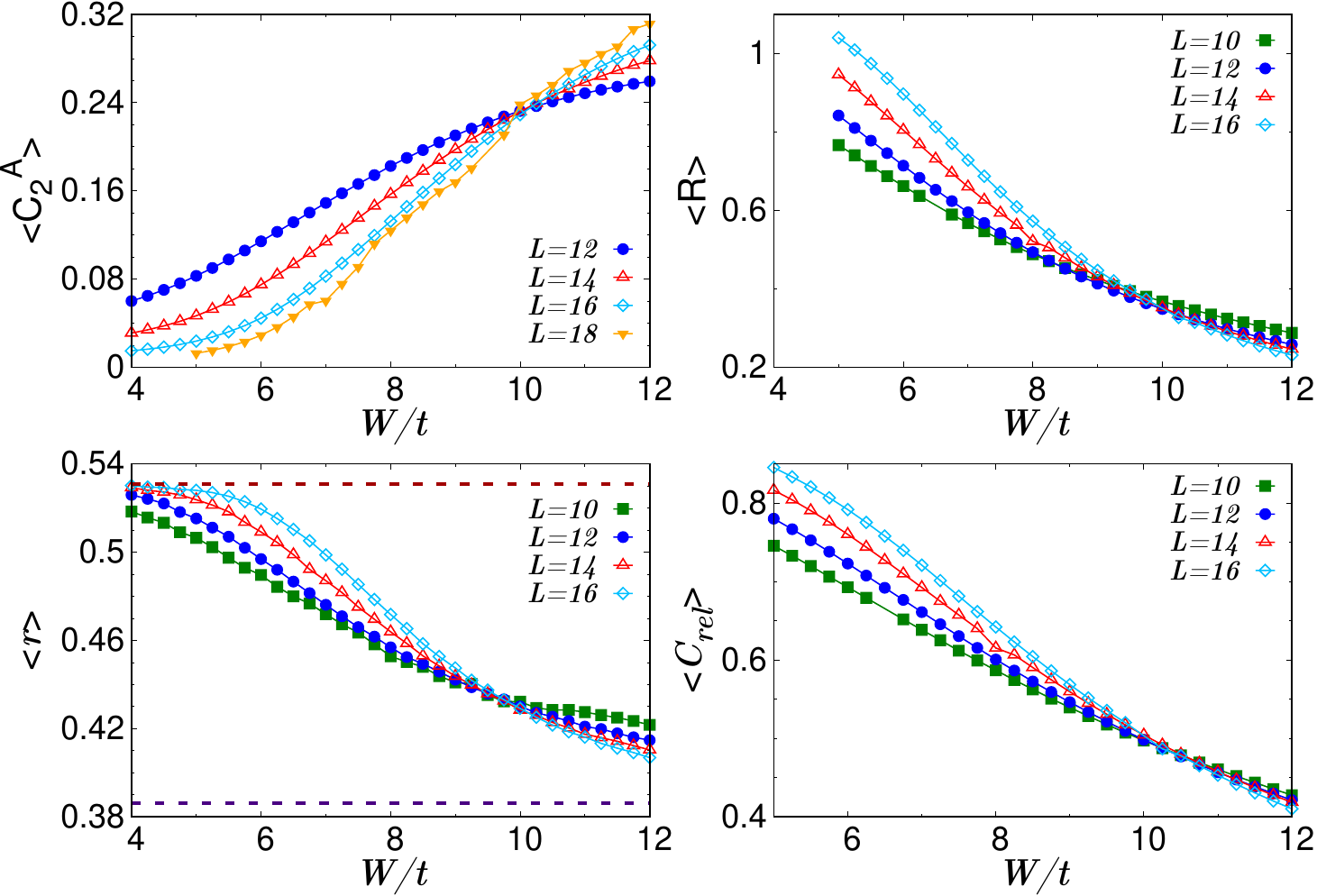}
  \caption{Top left panel shows the average value of the $l_2$ norm of sublattice quantum coherence, $\la C_2^{A} \ra$ as a function of disorder strength $W/t$ for various system sizes. $\la C_2^{A} \ra$ increases as the disorder strength increases. For weak disorder, $\la C_2^{A} \ra$ decreases with increase in $L$ while in the MBL phase $C_2^{A}$ increases with $L$ approaching one in the infinite size limit. Delocalization to MBL transition occurs around $W/t\sim 9.75t$. Top right panel shows bipartite EE $\la R \ra$ and the bottom left panel shows the average level spacing ratio $\la r \ra$ vs $W/t$.  Bottom right panel depicts the average value of the normalized relative entropy of coherence $\la C_{rel}\ra$  as a function of the disorder strength $W/t$. For weak disorder $\la C_{rel} \ra$ increases with $L$ approaching one while for $W \ge 10.0t$, $\la C_{rel}\ra$ decreases slowly with $L$ approaching zero in the MBL phase. All the quantities have been averaged over the entire eigen-spectrum as well as over ($50-5000$) independent disorder configurations for $L=18-10$ respectively.}
  \label{avg_fig}
\vskip-1cm
\end{center}
\end{figure}
Analysis of various trade-off relations between coherence and IPR in the previous section showed that coherence has very distinct behaviour for pure and mixed states in weakly and strongly disordered systems. Motivated by this, in this section we want to explore whether various norms of coherence can probe the delocalization to MBL transition. To be specific we focus on $l_2$ norm of subsystem coherence and relative entropy of coherence for subsystem as well as full system. 
Note that the $l_2$ norm of sublattice coherence is given by $C_2^A(E_n) = \sum_{\alpha \ne \beta} |\rho_{A,n}(\alpha\beta)|^2$. Top left panel of Fig.~\ref{avg_fig} shows $\la C_2^{A}\ra$ which is obtained by averaging $C_{2}^A(E_n)$ over the entire eigen spectrum as well as over a large number of independent disorder configurations.$C_2^A$ increases with increase in disorder $W$ incomplete contrast to the coherence for the full system. For weak disorder, in the delocalized phase, $\la C_2^A \ra$ decreases with increase in the system size such that for a large system $C_2^A$ is vanishingly small. In contrast to this in the presence of very strong disorder $\la C_2^A \ra$ increases as $L$ increases remaining always less than one. Delocalization to MBL transition occurs around $W\sim 9.75t$. 
 \begin{figure}[]
  \begin{center}
  \includegraphics[width=3.0in]{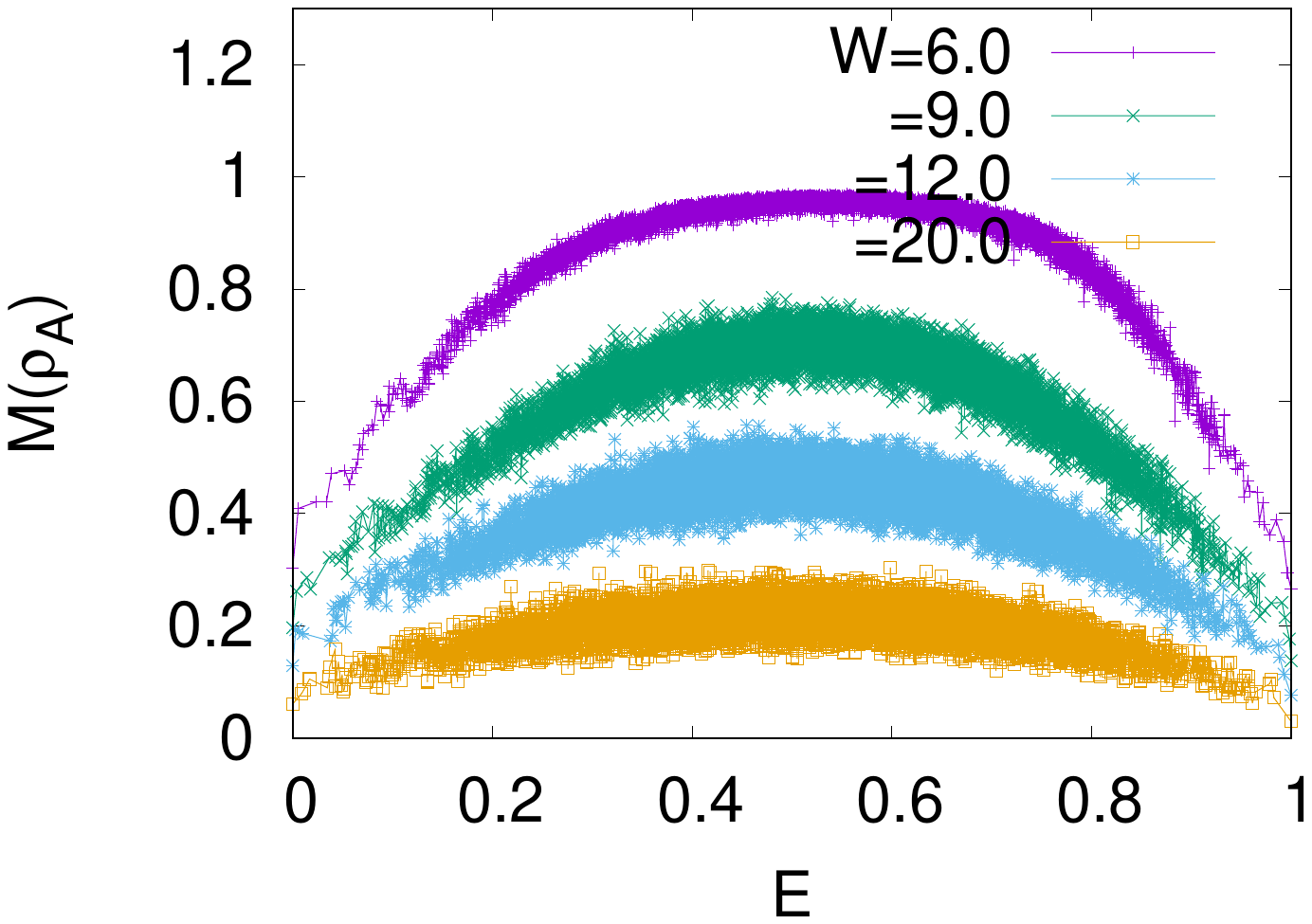}
  \caption{Mixedness of reduced density matrix $\rho_A$ vs rescaled energy $E$ for various disorder strengths. The data shown is for one disorder configuration.}
  \label{M}
  \end{center}
  \end{figure}
  Note that behaviour of sublattice coherence is completely opposite to the coherence for the full system which is largest for maximally extended state and goes to zero for a localized state. Our numerical results for $C_2^A$ are consistent with the behaviour of mixedness of the reduced density matrix $\rho^A$. Since $C_2^A+M^A \le 1$~\cite{Kumar} for any mixed state, and as shown in Fig.~\ref{M}, mixedness $M^A$ decreases with increase in the disorder strength $W$ for the model under consideration, $C_2^A$ increases with disorder.

\begin{figure*}[]
  \begin{center}
%    \vskip 0.5cm
   \hspace{-0.8cm}
  \includegraphics[width=3.3in]{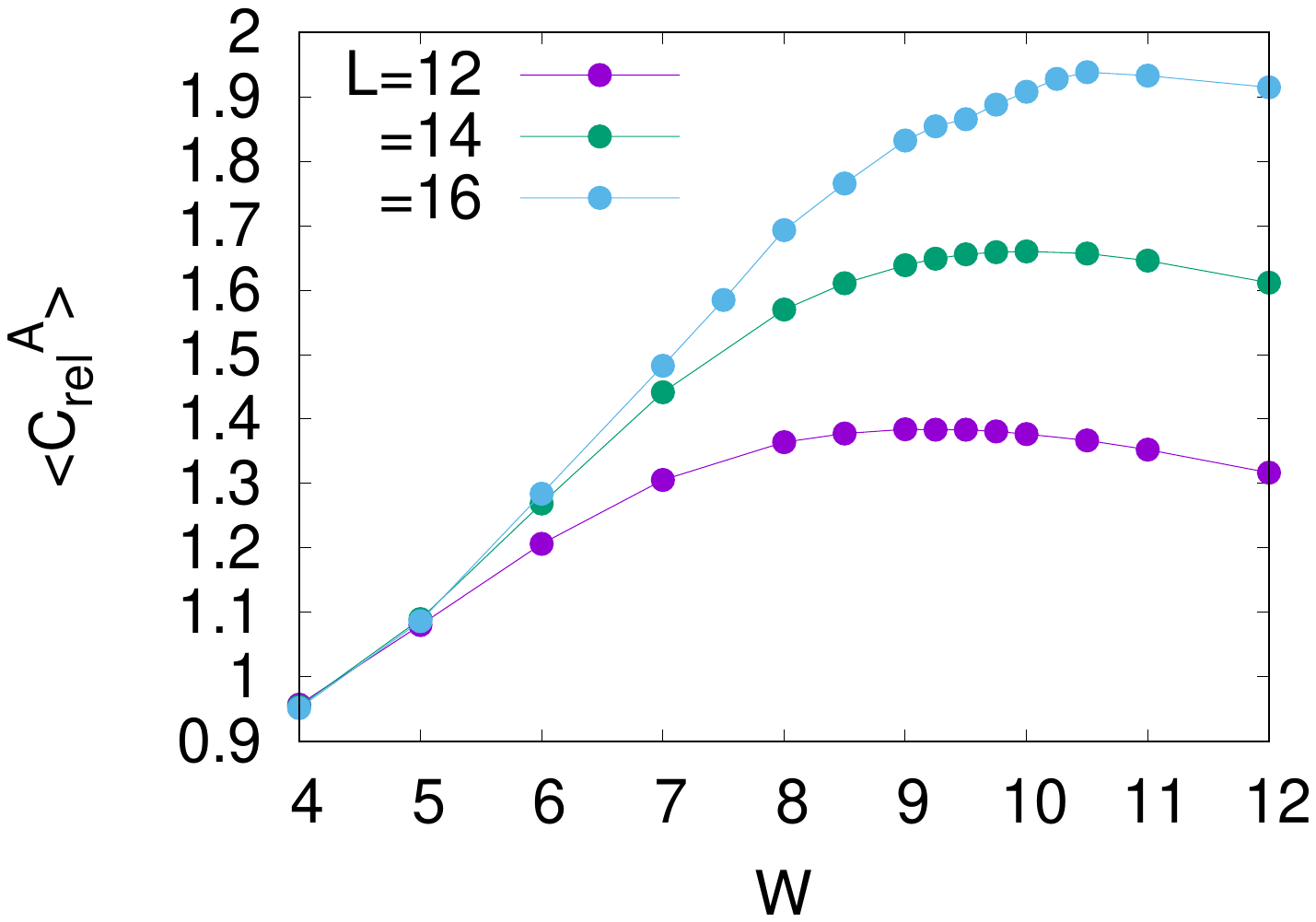}
  \hspace{-1.0cm}
    \includegraphics[width=3.3in]{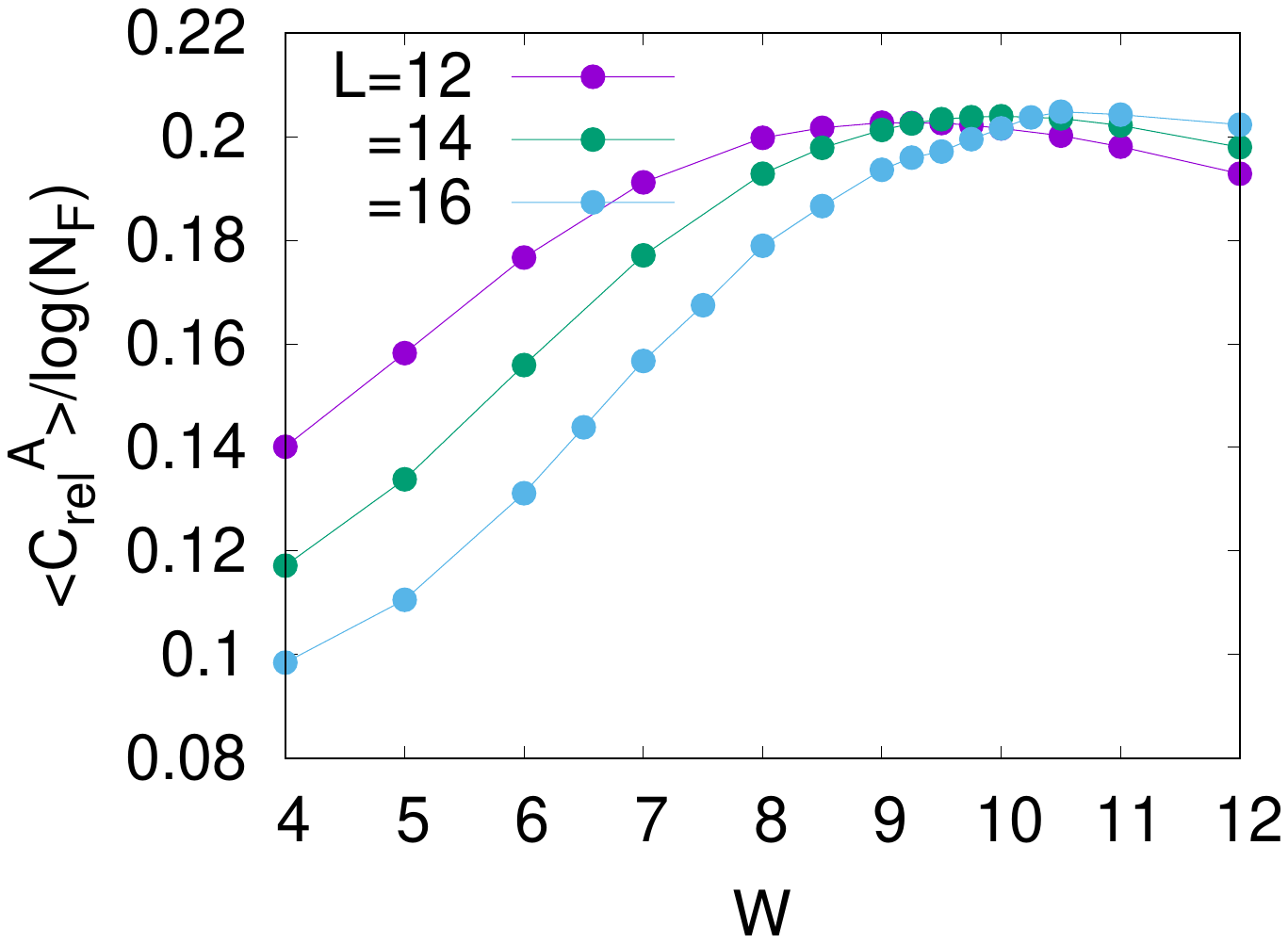}
  \caption{The left panel shows relative entropy of coherence for subsystem $A$ vs disorder $W$ for various system sizes. $C_{rel}^A$ increases as $W$ increases in contrast to $C_{rel}$. The right panel shows the plot of normalized relative entropy of coherence $C_{rel}^A/log(N_F)$ vs $W$ which shows a transition from delocalized phase to the MBL phase. The crossing point is very close to the one estimated from the level spacing ratio, bipartite entanglement entropy and $C_2^A$. The data shown is averaged over the eigenstates in the middle of the spectrum with $0.48 \le E \le 0.52$ and over many independent disorder configurations.}
\label{crelA}
\vskip-1cm
\end{center}
\end{figure*}

We also calculated the bipartite entanglement entropy (EE), $R(E_n)= -\log [Tr_A (\rho_A(E_n))^2]$.  %EE is expected to obey the volume law of scaling $S \sim L^d$ in the ergodic extended phase while it is suppressed for the MBL phase showing close to an area law scaling $S \sim L^{d-1}$ ~\cite{Huse,Sdsarma,Nayak,garg}. 
To minimize the finite-size effects we normalize the averaged $\la R(E_n) \ra$ with the value of bipartite EE within random-matrix theory, that is $R_{RMT}= L/2 \log(2)+[1/2+\log(1/2)]/2-1/2$~\cite{EE_RMT}. As shown in the top right panel of Fig.\ref{avg_fig}, the normalized and averaged $\la R \ra$ is larger when the sublattice coherence $\la C_2^A \ra$ is smaller and vice-versa. Our numerical observations are consistent with known relation between sublattice coherence and EE in general for bipartite quantum systems~\cite{coh_ee1}. The crossing point obtained from Renyi entropy, which is a well known characterization of the MBL phase, is close to the one obtained from the sublattice coherence.

We further confirm our findings about the transition point obtained from $\la C_2^A\ra$ by analysing another conventional characteristic of MBL transition, namely, the level spacing ratio. We calculate the ratio of successive gaps in energy levels $r_n=\frac{min(\delta_n,\delta_{n+1})}{max(\delta_n,\delta_{n+1})}$~\cite{Huse_2007} with $\delta_n=E_{n+1}-E_n$. The distribution of energy level spacing is expected to follow Poisson statistics with average value of $\la r \ra$ is $2\ln 2-1 \approx 0.386$ for localized phase while it follows Wigner-Dyson statistics with $ \la r\ra \approx 0.5295$ for the ergodic phase~\cite{Mehta1990}. As shown in the bottom left panel of Fig.~\ref{avg_fig}, average level spacing ratio, $\la r\ra$, also shows a transition around $W/t \sim 9.75$ in complete consistency with the transition point obtained from the sublattice coherence.  

We have explored the relative entropy of coherence $C_{rel}(E_n)$ of the full system as well as for the subsystem A for each eigenstate of the system under consideration. In the bottom right panel of Fig.~[\ref{avg_fig}], we have shown average relative entropy of coherence for the full system normalized by its maximum value of $log(N_F)$. $\la C_{rel} \ra$ decreases as the disorder strength increases. This is consistent with the trend of $C_1$ in disordered systems~\cite{coh_MBL1}. For weak disorder, $\la C_{rel}\ra$ increases with $L$ approaching one while in the localized phase $\la C_{rel}\ra $ decreases with $L$ slightly. A transition is observed at around $W_c \sim 10.0t$ which is close to the transition obtained from $l_2$ norm of sublattice coherence and bipartite entanglement entropy. Normalized values of quantum coherence $C_1/(N_F-1)$ and the relative entropy of coherence $C_{rel}/\log N_F$ are of order one for highly delocalized states and approach zero for a highly localized state. Thus,  a maximally extended state is also maximally coherent. 

Next, we analyse the relative entropy of coherence for the subsystem A. Fig.~\ref{crelA} shows $C_{rel}^A$ as a function of disorder for various system sizes obtained by doing aveaging over the eigenstates in the middle of spectrum with $0.48 \le E \le 0.52$ where $E$ is the rescaled energy $E=\frac{E_n-E_{min}}{E_{max}-E_{min}}$. As $W$ increases, $C_{rel}^A$ increases in complete contrast to $C_{rel}$ for the full system shown in the right bottom panel of Fig.~\ref{avg_fig}. Since in the limit of very strong disorder, diagonal entropy $S(\rho_D)$ as well as bipartite von-Neumann entropy $S$ would become vanishingly small, there is a downturn in $C_{rel}^A$ beyond the transition point for $W \gg W_c$. Normalised value of relative entropy of sublattice coherence $C_{rel}^A/log(N_F)$ captures the delocalization to MBL transition point.  
Thus, in an interacting disordered quantum system, $C_2^{A}$ and $C_{rel}^A$, both of which are related to the notion of localizable coherence~\cite{Hamma}, increase with increase in disorder in contrast to the coherence for the full system discussed before. 

At this end, we would like to mention that the system size dependence of the transition point determined from various norms of coherence seems to be slightly weaker than that for other diagnostics of MBL like spectral form factor~\cite{Prosen}, entanglement entropy and level spacing ratio which are known to show logarithmic dependence on $L$ of the transition point~\cite{Piotr_AA}. Furthermore, though our finite-size data seems to show a finite transition point to MBL phase like all earlier works, we can not rule out the possibility of diverging resonances and quantum avalanche leading to unstable MBL phase in the infinite size limit as was pointed out earlier~\cite{avalanche}. Though it has also been shown that the avalanche for static MBL systems like the one we discussed here may actually terminate leading to a stable MBL phase beyond a finite disorder even in the infinite size limit.~\cite{Piotr_avalanche}. Nonetheless, enhanced subsystem coherence in strong disorder regime of a finite disordered interacting system is of relevance from the point of view of quantum devices. 

We have also studied these quantities as a function of energy eigenvalues for a fixed disorder strength and observed consistency with earlier works on MBL~\cite{Alet_rev,Vidmar_rev,garg} details of which are provided in the next section.

\section{Energy resolved sublattice quantum coherence}
It is also interesting to explore the sublattice coherence $C_2^A$ and $C_{rel}^A$  as a function of many-body energy eigen-values to look for crossover from delocalization to MBL for a fixed disorder strength. 
\begin{figure}
  \begin{center}
    \vskip0.5cm
 %   \hspace{-1cm}
  \includegraphics[width=3.4in]{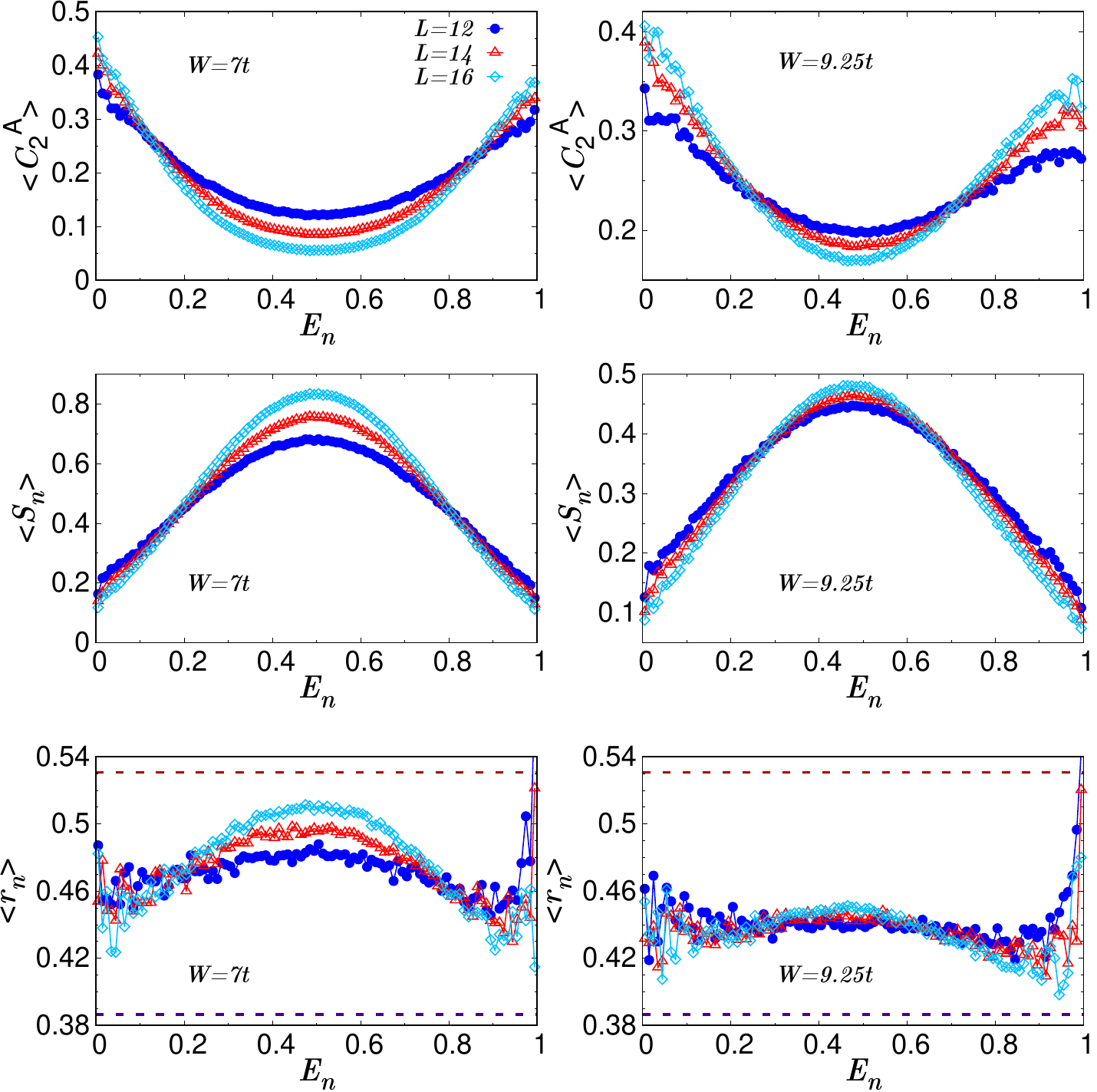}
  \caption{Top row shows the $l_2$-norm of sublattice coherence $\la C_2^A \ra$ vs rescaled energy $E$ for $W=7t$ and $W=9.25t$ for three $L$ values. For eigenstates in the middle of the spectrum $\la C_2^A \ra$ decreases as $L$ increases indicating that these are extended states and for eigenstates on the edges of the spectrum $\la C_2^A \ra$ slightly increases with $L$. The width of the $E$ region for which $\la C_2^A \ra$ decreases with $L$ is smaller for $W=9.25t$ which is close to the transition point. Middle panel shows bipartite entanglement entropy vs $E$ and the third row shows level spacing ratio vs $E$. All the quantities presented here have been averaged over a large number of independent disorder configurations.}  
  \label{fig2}
\vskip-1cm
\end{center}
\end{figure}
Here, in Fig.~[\ref{fig2}] we present disorder averaged $\la C_2^A(E) \ra$ vs the rescaled  eigen energy $E=\frac{E_n - E_{min}}{E_{max}-E_{min}}$ for three system sizes. For $W =7.0t <W_c$, for eigen-states in the middle of the spectrum with $0.175 \le E \le 0.85$ $\la C_2^A(E)\ra$ decreases as the system size increases while for $E>0.85$ and $E <0.175$, $\la C_2^A(E)\ra$ increases with the system size approaching one. As the disorder strength increases, the range of eigenvalues in the middle of the spectrum, for which states are extended, decreases as the strength of disorder increases. This is depicted in the top right panel of Fig.~[\ref{fig2}] which presents $\la C_2^A(E) \ra$ for a value of disorder $W=9.25t$. For $W =9.25t$, which is still on the delocalized side of the transition point, a much smaller fraction of eigenstates in the middle of the spectrum $0.35 \le E \le 0.65$ show decrease in $\la C_2^A(E)\ra$ as $L$ increases. This trend is consistent with what is observed in well-known characterizations of the MBL, namely, bipartite entanglement entropy (EE) and level-spacing ratio as shown in the second and the third row of Fig.~[\ref{fig2}].

Bipartite entanglement entropy increases with system size approaching the Page value for states in the middle of the spectrum while it is almost independent of the system size for states on the edges of the spectrum. 
As shown in Fig~[\ref{fig2}], the range of eigenvalues for which $\la R_n\ra$ has very weak system size dependence increases with the disorder strength. This also indicates that eigenstates with larger $\la R_n\ra$ have lower sublattice coherence. 
As shown in the third row of Fig.~\ref{fig2}, level spacing ratio $\la r_n \ra$ increases with $L$, approaching the average for Wigner-Dyson statistics for eigen-states in the middle of the spectrum. In contrast to this for eigenstates on the edges of the spectrum, level spacing ratio decreases with $L$ approaching the Poissonian value. 
The range of eigenvalues showing Wigner-Dyson statistics is consistent with the range of eigenvalues for which $\la C_2^A(E)\ra$ decreases with the system size. 
These observations are consistent with earlier works on MBL~\cite{Alet_rev,Vidmar_rev,garg}.  %This analysis demonstrates that $l_2$ norm of sublattice coherence $C_2^A$ is also an ideal physical quantity to characterize the delocalization to MBL transition. 
\begin{figure}[ht]
  \begin{center}
%    \vskip0.5cm
    \hspace{-0.8cm}
  \includegraphics[width=2.0in]{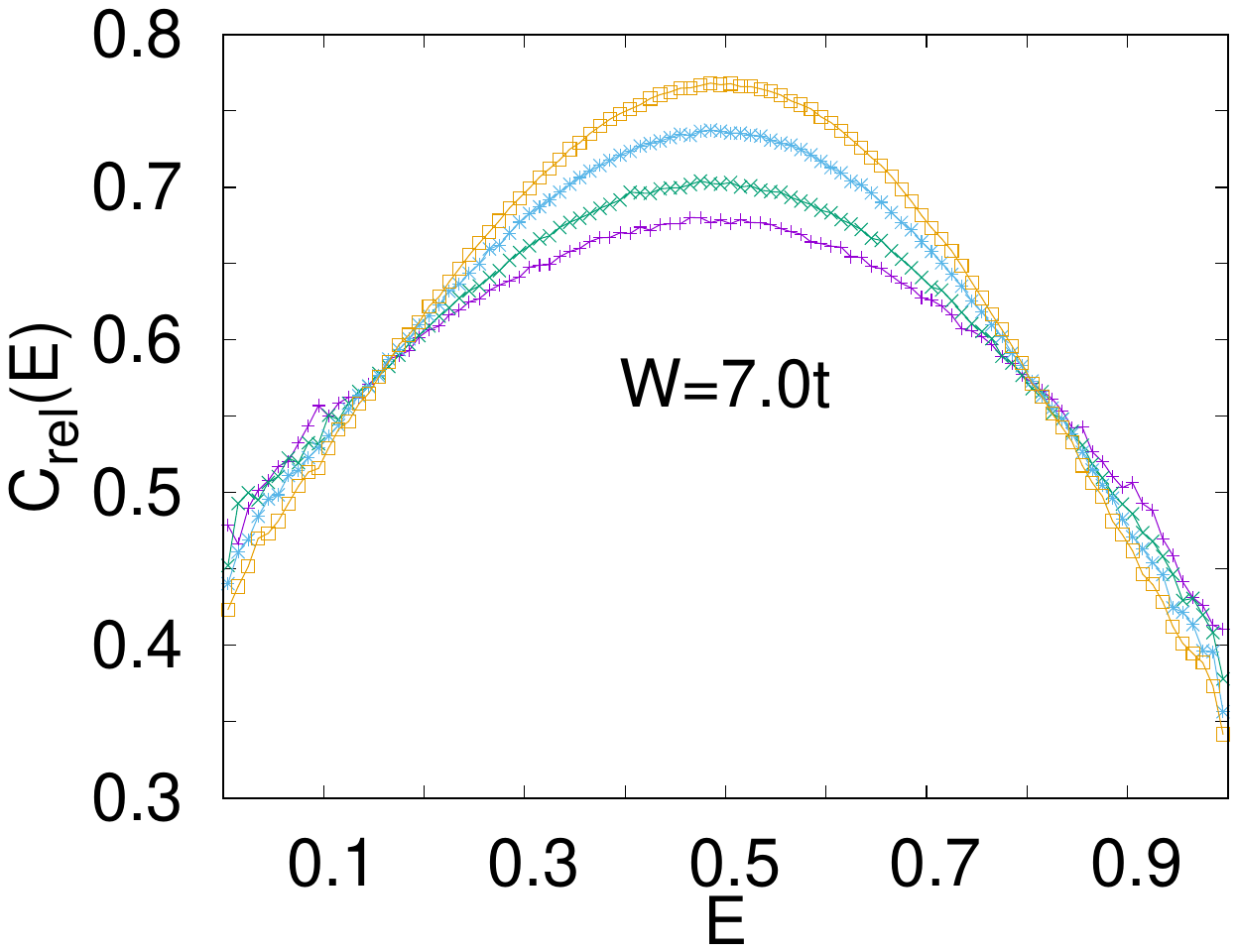}
%  \vskip-4cm
  \hspace{-1.0cm}
  \includegraphics[width=2.0in]{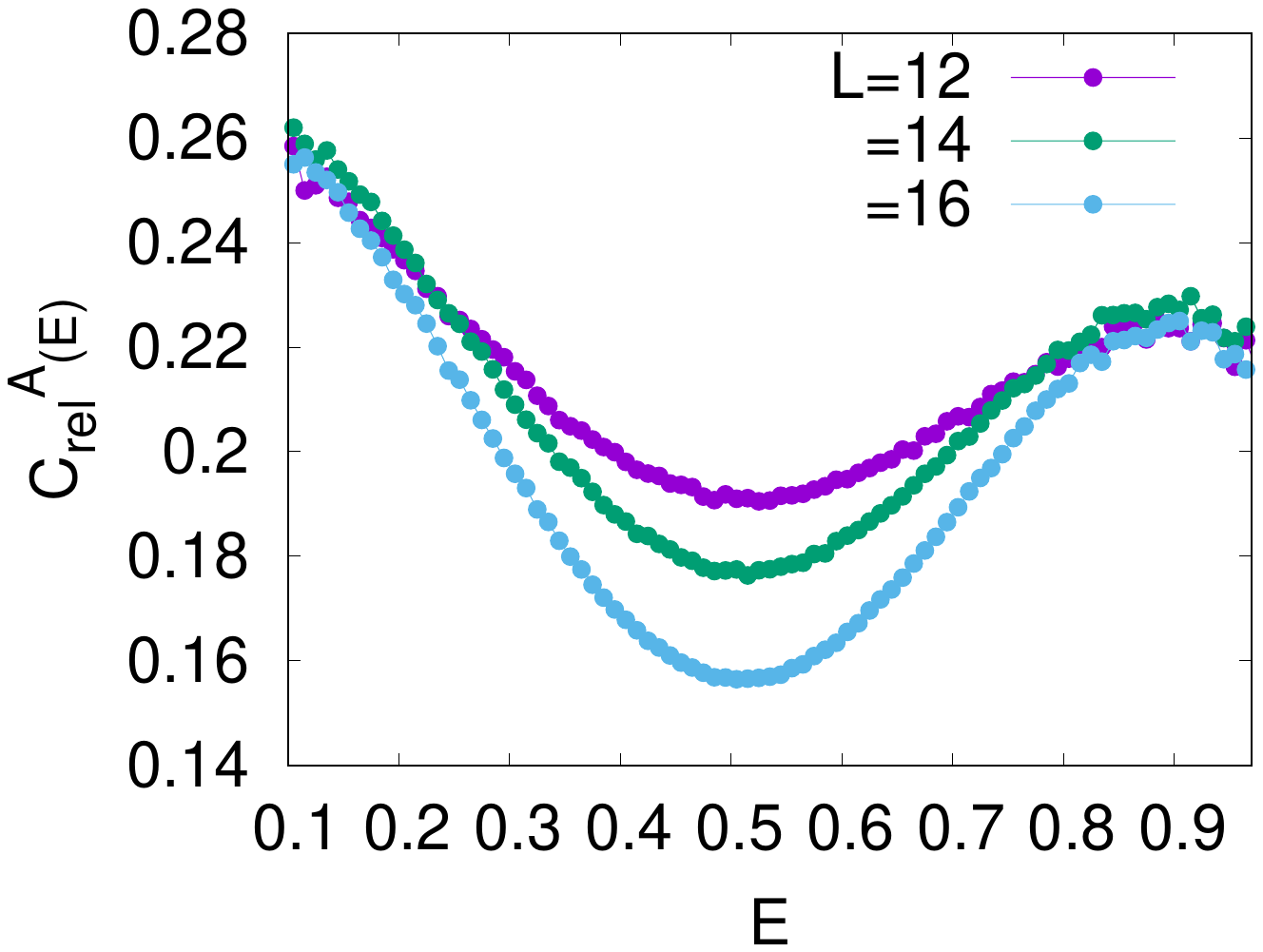}
  \caption{Normalized relative entropy of coherence $C_{rel}(E)$ and sublattice coherence $C_{rel}^A(E)$ vs rescaled eigen energy $E$ for $W = 7.0t$ for $L=10,12,14,16$ from bottom to top. For states in the middle of the spectrum $C_{rel}(E)$ increases as $L$ increases while $C_{rel}^A$ decreases but for eigenstates on the edges of the spectrum $C_{rel}(E)$ decreases while $C_{rel}^A$ increases.}  
  \label{fig3}
\vskip-1cm
\end{center}
\end{figure}
Based on energy dependence of various physical quantities it is obvious to think that states in the middle of the spectrum should be analysed to find the disorder strength at which delocalization to MBL transition takes place.
    As the density of many-body states is peaked in the middle of the spectrum, an ensemble average over the entire spectrum with equal weight for each eigenstate is almost equivalent to average over eigenstates the middle of the spectrum ~\cite{Subroto}. This is clear from the disorder dependence of mid-spectrum averaged level spacing ratio and $C_2^A$ discussed in Appendix A. 
    
Fig.~[\ref{fig3}] shows normalized relative entropy of coherence $C_{rel}(E)$ and sublattice relative entropy of coherence $C_{rel}^A(E)$ averaged over many independent disorder configurations, versus rescaled energy eigenvalues $E$.  There is an increase in $C_{rel}(E)$ with chain size, $L$, for eigenstates in the middle of the spectrum while $C_{rel}^A(E)$ decreases with $L$ for eigenstates in the middle of the spectrum. 
Furthermore, the subsystem $C_{rel}^A(E)$ remains finite with almost no significant system-size dependence for states at the top and bottom of the spectrum. 
This implies that just like $l_2$ norm of sublattice coherence,  sublattice relative entropy of coherence is also an important characterization of MBL systems. More importantly, this is significant from the point of view of quantum processors and other quantum devices like superconducting qubit arrays. Our analysis suggests that some sources of randomness in the qubit array can help in enhancing coherence for a subsystem of qubit array for at least low energy states even when disorder is not very large.

\section {Conclusions}
Quantum coherence is a distinguishing property of quantum mechanics. The exact relations between quantum coherence and a measure of localization is a signature of the fundamental role of quantum mechanics and quantum coherence in the physics of localization. Tradeoff relations derived in this work between IPR and coherence imply that various measures of quantum coherence are ideal characterizations for the delocalization to many-body localization transition. 
The relations between IPR and the relative entropy of coherence also explain why bipartite EE carries signatures of the localization-to-delocalization transition. Enhanced sublattice coherence leads to lower bipartite EE, and vice versa. 
%Relations derived here are generic and hold true for any disorder strength for both the interacting as well as non-interacting systems though we have shown their applications only for the case of delocalization to MBL transition. 
 Some interesting observations from this study are that, though in an MBL phase coherence of the full system is lost, the subsystem coherence is retained and is large.  In other words coherence of a pure random eigenstate is suppressed by MBL but MBL may help in enhancing the coherence of a mixed state obtained from reduced density matrix of a subsystem. A simple intuitive explanation for this observation can be provided in terms of non-ergodic nature of the MBL phase. Generally a system looses its coherence due to coupling with the environment. In an isolated system in the MBL phase, a part of the system fails to thermalize with the rest of the system acting as its bath because of extremely weak entanglement between subsystem A and the rest of the system. Interestingly, this weak entanglement also helps a subsystem to preserve its coherence in the MBL phase. 
  
 %In complete contrast to this, a system in MBL phase has small coherence for the whole system but a large sublattice coherence.  
 
Quantum computation and communication rely heavily on entanglement and coherence. Preparing multi-qubit entangled states is crucial for optimal performance of quantum computers.
%~\cite{quan_comp} , while quantum coherence is a key resource for creating'magic' ~\cite{Pati_magic}.
%Quantum engineers have significant challenges in controlling entanglement and coherence. 
Our findings on the relation between coherence and localisation are significant as they suggest a new approach to controlling quantum coherence by manipulating inhomogeneities and interactions in a system. For example, in superconducting qubit arrays, Josephson energies can be tuned to govern localisation, quantum coherence, and entanglement~\cite{MBL_qubits}. Our analysis shows that if Josephson energies are inhomogeneous in a superconducting qubit array, coherence can be enhanced for a subsystem of qubits due to proximity of an MBL phase. 
Finally, from a theoretical perspective, investigating these tradeoff relations helps in developing a more comprehensive framework for quantum thermodynamics and quantum statistical mechanics. It opens new avenues and pushes the boundaries of how we understand information flow and state preservation in complex quantum systems. As we move towards building larger and more sophisticated quantum devices, understanding these tradeoffs will be crucial in guiding the design of robust and scalable quantum technologies.

%Our findings suggest a need for more research on the relationship between coherence and many-body localisation.
%Entanglement and quantum coherence play a pivotal role in the field of quantum computation and quantum communication, for example, it is essential to prepare a multi-qubit entangled states for optimal functioning of quantum computers~\cite{quan_comp} and quantum coherence is the fundamental resource for creation of 'magic'~\cite{Pati_magic}. Thus, how to control entanglement and coherence in a quantum system are among the biggest challenges for quantum engineers . In this context the exact relations derived between measure of coherence and localization in this work are significant because they suggest a novel way to control the quantum coherence by playing with the inhomogeneities and interactions in the system to control the measure of many-body localization and hence coherence. For example in a superconducting qubit array, one can tune the inhomogeneities of Josephson energies to control the measure of localization, quantum coherence and entanglement~\cite{MBL_qubits}.  We believe the concepts developed in this work would encourage more detailed investigations into the link between coherence and many-body localization.    
\section{Acknowledgements}
A.G. would like to acknowledge discussions with F. Alet and N. Laflorencie at Info-French discussion meeting ICTS/IFWCQM2024/12 held at ICTS. A. G would also like to acknowledge Y. Prasad  for help with figures.
\section{Appendix A: Mid-spectra averages}
 \begin{figure*}
  \begin{center}
  \includegraphics[width=3.0in,angle=0]{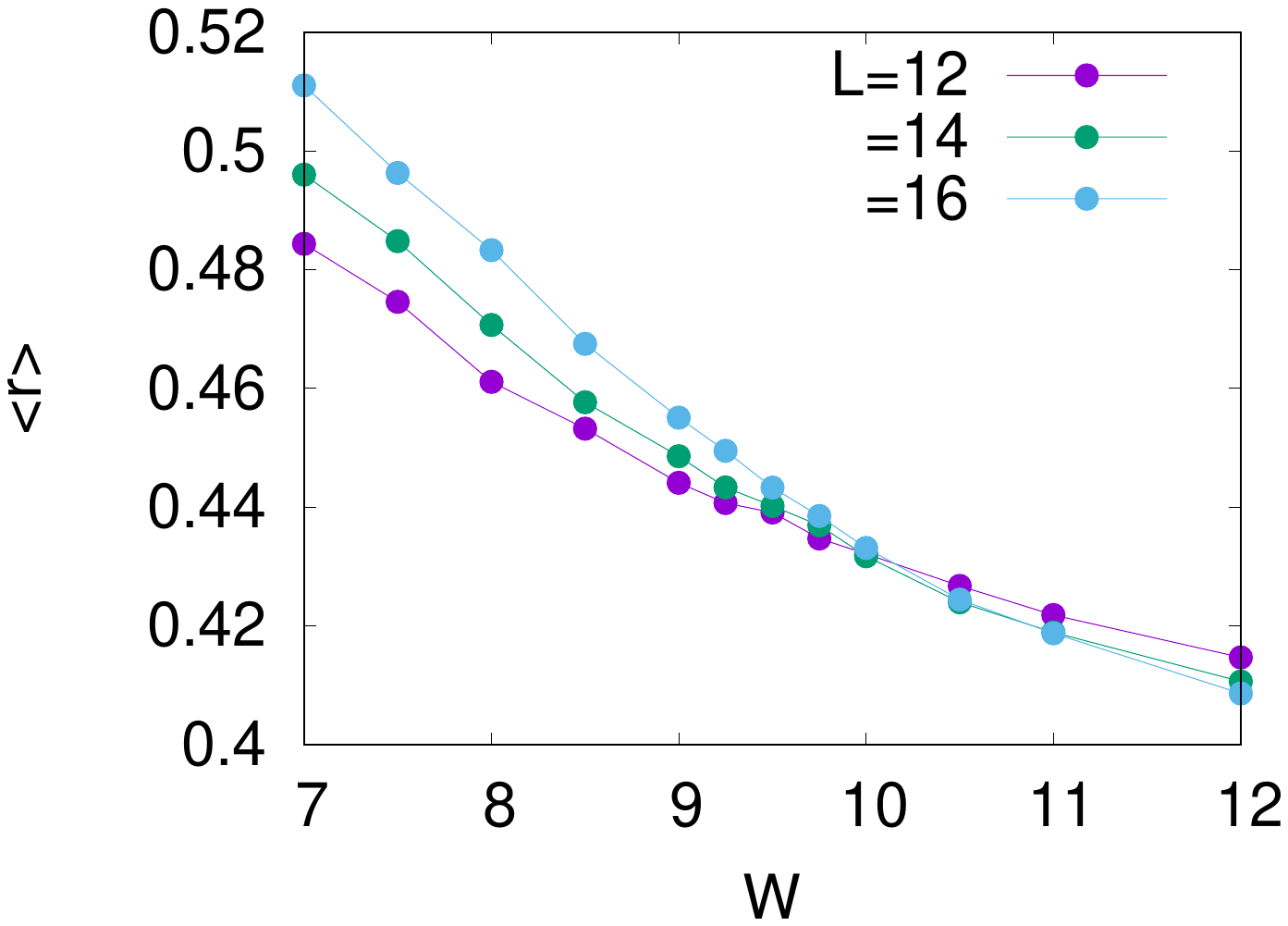}
  \includegraphics[width=3.0in,angle=0]{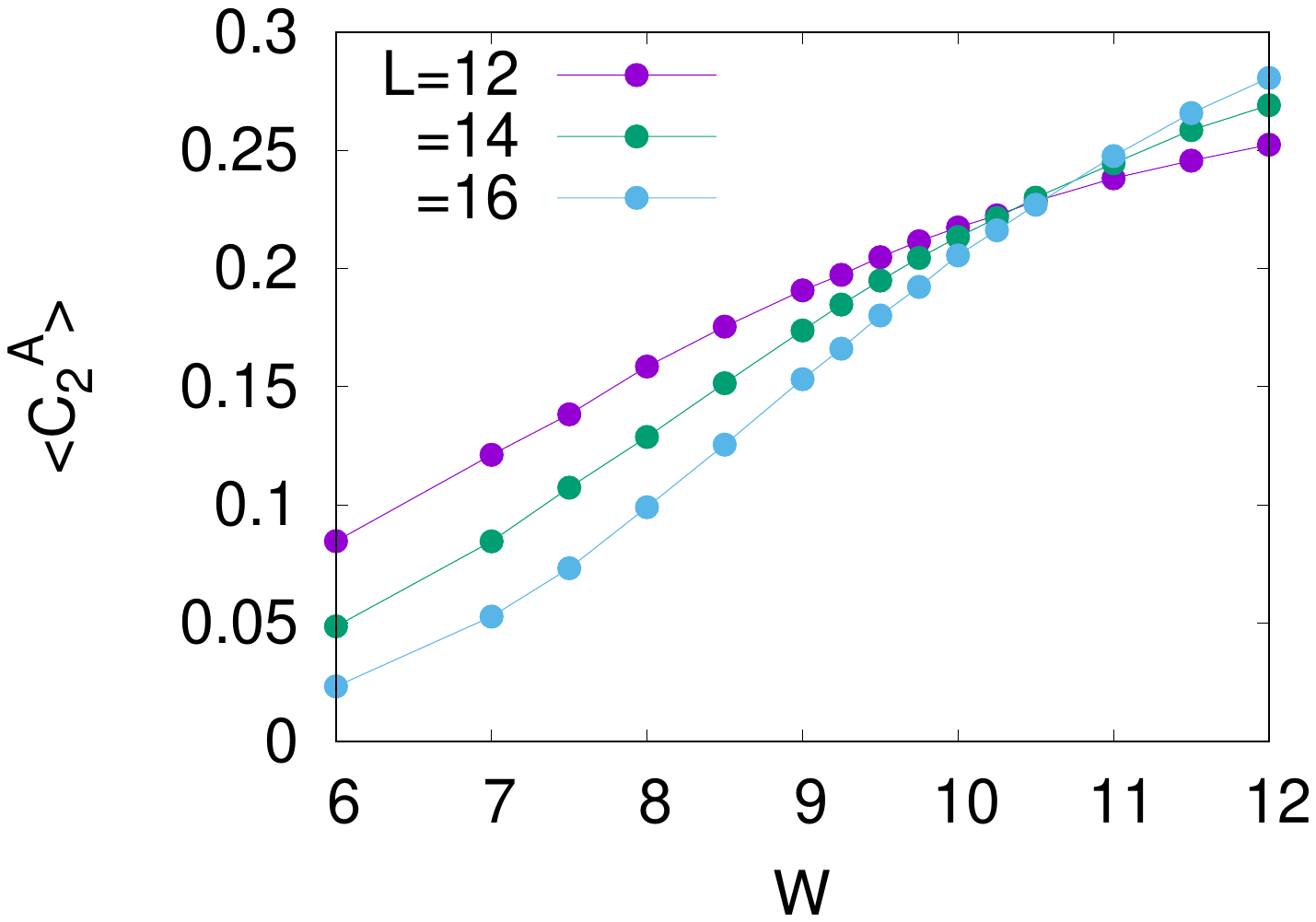}
  \caption{Level spacing ratio, and $C_2^A$ vs $W$ obtained from averaging over middle eigenstates ($0.48 \le E \le 0.52)$ where $E$ is the rescaled eigenvalue normalized between 0 to 1. The transition point obtained from mid-spectra average is almost the same as the one obtained from average over the entire spectrum.}
  \label{mid_av}
  \end{center}
   \end{figure*}
 In the main paper, we presented the $l_2$ norm of subsystem coherence $C_2^A$ and the level spacing ratio averaged over the entire eigen-spectrum. But the energy resolved quantities shown in Fig.~[\ref{fig2},\ref{fig3}] indicate that the system probably has a mobility edge. There may be a crossover from localized to delocalized states as one moves in the energy-eigen spectrum for a fixed strength of disorder such that eigenstates in the middle of the spectrum get localized at a larger $W$ value than the states at the edges of the spectrum. This is in consistency with most of the earlier works on MBL.  Thus, it is meaningful to explore the transition point by analyzing various physical quantities obtained only for eigenstates in the middle of the spectra. Here we present results obtained by averaging contributions only from the  eigenstates in the middle of the spectrum. For a rescaled energy $E=\frac{E_n-E_{min}}{E_{max}-E_{min}}$, we average various physical quantities for eigenstates in the middle of the spectrum for $0.48 \le E \le 0.52$.  As shown in Fig.~[\ref{mid_av}], both the level spacing ratio  and the $l_2$ norm of subsystem coherence show a transition point which is consistent with the one obtained from the average over the entire spectrum within numerical precision.  
 Since density of many-body states peak in the middle of the spectrum, an ensemble average over the entire spectrum with equal weight for each eigenstate is almost equivalent to the average only in the middle of the spectrum ~\cite{Subroto}.  
 %which is consistent with the infinite temperature ensemble averages presented in the main paper.
    \\\\

\bibliography{bibliography}

\end{document}